\documentclass{ecai} 
\usepackage{latexsym}
\usepackage{amssymb}
\usepackage{amsmath}
\usepackage{amsthm}
\usepackage{booktabs}
\usepackage{enumitem}
\usepackage{graphicx}
\usepackage{color}
\usepackage{algorithm}
\usepackage{algpseudocode}
\usepackage[switch]{lineno}
\newtheorem{theorem}{Theorem}

\newcommand{\BibTeX}{B\kern-.05em{\sc i\kern-.025em b}\kern-.08em\TeX}

\begin{document}

%%%%%%%%%%%%%%%%%%%%%%%%%%%%%%%%%%%%%%%%%%%%%%%%%%%%%%%%%%%%%%%%%%%%%%%%

\begin{frontmatter}

%%% Use this command to specify your submission number.
%%% In doubleblind mode, it will be printed on the first page.

\paperid{7632} 

%%% Use this command to specify the title of your paper.

\title{ToMacVF : Temporal Macro-action Value Factorization for Asynchronous Multi-Agent Reinforcement Learning}

%%% Use this combinations of commands to specify all authors of your 
%%% paper. Use \fnms{} and \snm{} to indicate everyone's first names 
%%% and surname. This will help the publisher with indexing the 
%%% proceedings. Please use a reasonable approximation in case your 
%%% name does not neatly split into "first names" and "surname".
%%% Specifying your ORCID digital identifier is optional. 
%%% Use the \thanks{} command to indicate one or more corresponding 
%%% authors and their email address(es). If so desired, you can specify
%%% author contributions using the \footnote{} command.

\author[A]{\fnms{Wenjing}~\snm{ZHANG}\orcid{0009-0006-2571-604X}}
\author[A]{ \fnms{Wei}
~\snm{ZHANG} \orcid{0000-0002-0598-4606}\thanks{Corresponding Author. Email: weizhang@hit.edu.cn}}
% \author[B]{}

\address[A]{Harbin Institute of Technology}

%%% Use this environment to include an abstract of your paper.

\begin{abstract}
Existing asynchronous MARL methods based on MacDec-POMDP typically construct training trajectory buffers by simply sampling limited and biased data at the endpoints of macro-actions, and directly apply conventional MARL methods on the buffers.
As a result, these methods lead to an incomplete and inaccurate representation of the macro-action execution process, along with unsuitable credit assignments.
To solve these problems, the {\bf T}emp{\bf o}ral {\bf Mac}ro-action {\bf V}alue {\bf F}actorization (ToMacVF) is proposed to achieve fine-grained temporal credit assignment for macro-action contributions.
A centralized training buffer, called {\bf Mac}ro-action {\bf S}egmented {\bf J}oint {\bf E}xperience {\bf R}eplay {\bf T}rajectory (Mac-SJERT), is designed to incorporate with ToMacVF to collect accurate and complete macro-action execution information, supporting a more comprehensive and precise representation of the macro-action process.
To ensure principled and fine-grained asynchronous value factorization, the consistency requirement between joint and individual macro-action selection called {\bf T}emp{\bf o}ral {\bf Mac}ro-action based {\bf IGM} (To-Mac-IGM) is formalized, proving that it generalizes the synchronous cases.
Based on To-Mac-IGM, a modularized ToMacVF architecture, which satisfies CTDE principle, is designed to conveniently integrate previous value factorization methods.
Next, the ToMacVF algorithm is devised as an implementation of the ToMacVF architecture.
Experimental results demonstrate that, compared to asynchronous baselines, our ToMacVF algorithm not only achieves optimal performance but also exhibits strong adaptability and robustness across various asynchronous multi-agent experimental scenarios.
\end{abstract}

\end{frontmatter}

%%%%%%%%%%%%%%%%%%%%%%%%%%%%%%%%%%%%%%%%%%%%%%%%%%%%%%%%%%%%%%%%%%%%%%%%
\section{Introduction}
% 第一段简要介绍MARL和值分解的发展现状
Multi-agent reinforcement learning(MARL) has made remarkable achievements in various cooperative domains, including traffic signal control \cite{traffic1,traffic2}, robot collaboration \cite{robot,robot2}, and simulation game AI \cite{game,game2}.
Due to partial observability and communication constraints, cooperative MARL methods based on decentralized partially observable Markov decision processes(Dec-POMDP) \cite{dec-pomdp} typically adopt a centralized training with decentralized execution (CTDE) \cite{ctde} architecture to facilitate effective collaboration.
Previous value factorization methods \cite{qmix,qtran,qplex} based on CTDE decompose the joint value function derived from global rewards into the local individual utilities of each agent, thereby promoting efficient end-to-end training of decentralized policies.
These methods adhere to the Individual Global Max (IGM) principle  \cite{qtran} to ensure consistency between local greedy action selection during decentralized execution and joint greedy action selection during centralized training. 
Such approaches demonstrate impressive performance across various aspects, including improving the expressiveness of the mixing network \cite{qplex}, preventing sub-optimal convergence \cite{wqmix,suboptimal}, promoting exploration \cite{maven,uneven}, and integrating diverse functional modules \cite{qatten,smix,implicit}.

% 说缺点1：同步值分解方法不适用于异步场景
Despite the successes of previous value factorization methods, significant limitations remain when addressing asynchronous MARL scenarios.
In many real-world scenarios that require asynchronous decision-making, agents execute macro-actions with inconsistent durations and initiate or terminate these actions at different timesteps.
However, previous MARL approaches assume synchronous training and execution, where all agents perform primitive actions simultaneously, each lasting only a single timestep.
This assumption makes them unsuitable for asynchronous scenarios.
%说缺点2：先前的集中式训练的异步方法（包括值分解方法）存在对于宏动作执行过程信息采样不完整，不准确的问题。
Subsequently, asynchronous MARL methods \cite{asy_marl1,asy_marl2,warehouse1,warehouse2,asy_marl3} based on macro-action decentralized partially observable Markov decision process(MacDec-POMDP) \cite{macPOMDP1,macPOMDP2} have been proposed. 
Asynchronous MARL methods with centralized training utilize a buffer called macro-action joint experience replay trajectories (Mac-JERTs) \cite{warehouse1}.
Mac-JERTs aims to treat macro-actions as primitive actions by sampling limited and biased information only at the macro-action termination timesteps.
Therefore, previous asynchronous MARL methods can train macro-action policies by idealistically applying conventional MARL methods designed for synchronous decision-making.
However, they neglect the temporal fine-grained information inherent in the execution process of macro-actions.
% Specifically, Mac-JERTs encounter the problems of duplicated macro-actions, inconsistent macro-observations\cite{reduntant}, and incorrect cumulative rewards.
As a result, these asynchronous MARL methods incorporated with Mac-JERTs produce incomplete and inaccurate representations of the macro-action execution process, along with incorrect estimation of macro-action values or probabilities. 
%说值分解方法也使用了先前方法的设置的缺点
Recently, asynchronous value factorization(AVF) methods have been proposed \cite{avf}, which also adopt the Mac-JERTs buffer settings. 
These methods integrate previous value factorization methods to train asynchronous macro-action policies in a synchronous manner, leading to unsuitable and incorrect credit assignments to the individual macro-action value functions over time.

In contrast, our method completely overturns the previous asynchronous MARL, which only adapted synchronous techniques to asynchronous settings without modification.
% 我们的方法
To address these limitations of previous AVF methods, the Temporal Macro-action Value Factorization(ToMacVF) framework is proposed to achieve fine-grained temporal credit assignment for macro-action contributions.
% Mac-SJERT
A fundamentally different centralized training buffer, called Macro-Action Segmented Joint Experience Replay Trajectory (Mac-SJERT), is incorporated with ToMacVF framework to ensure complete and accurate collection of macro-action execution information, thereby supporting more comprehensive and precise representations of the macro-action execution process.
%positional encoding 
% To solve the issue of duplicated macro-actions and inconsistent macro-observations, sinusoidal positional encoding\cite{Positional} is applied to incorporate temporal information into both macro-actions and macro-observations and ensure the collection of the real macro-observations at each timestep.
%ToMacVF
Moreover, to ensure principled and fine-grained asynchronous value factorization, the consistency requirement between joint and individual macro-action greedy selection called Temporal Macro-action-based IGM (To-Mac-IGM) is formalized, demonstrating that it can be generalized to IGM \cite{qplex}, which is used in synchronous cases.
Based on the To-Mac-IGM, the Agent-oriented Temporal Parameter Generation (ATPG) module is incorporated in ToMacVF to satisfy the sufficiency of To-Mac-IGM by generating non-negative parameters (weights and biases of the mixing network) for credit assignment across both time and agent dimensions.
As the ATPG module incorporates temporal inter-agent dependencies for asynchronous value factorization, it enhances effective inter-agent coordination.

%实验结果
To follow prior asynchronous MARL works and enable comparison, we conducted experiments on three open-source standard benchmarks for MacDec-POMDP: BoxPushing \cite{boxpushing}, Overcooked \cite{overcooked}, and WareHouse \cite{warehouse1}.
The experimental results demonstrate that, compared to state-of-the-art asynchronous MARL baselines, our ToMacVF not only achieves optimal performance but also exhibits strong adaptability and robustness across various asynchronous multi-agent experimental scenarios.
Furthermore, ablation studies validate the effectiveness of each proposed module.
In summary, the main contributions are as follows:
% 列举文章重点
\begin{itemize}
\item 
We propose a fundamentally different centralized training buffer, Mac-SJERT, which is incorporated in ToMacVF to ensure the complete and accurate collection of macro-action execution information, thereby enabling more comprehensive and precise representations of the macro-action execution process.
\item 
To ensure principled and fine-grained asynchronous value factorization, we formalize the consistency requirement between joint and individual macro-action greedy selection, named To-Mac-IGM. 
Furthermore, we demonstrate that To-Mac-IGM can be generalized to IGM, which is used in synchronous cases.
\item 
Based on To-Mac-IGM, our framework is equipped with the Agent-oriented Temporal Parameter Generation(ATPG) module to satisfy the sufficiency of To-Mac-IGM by generating non-negative parameters.
Besides, our method can conveniently integrate previous value factorization methods.
\item 
The experimental results demonstrate that, compared to state-of-the-art asynchronous MARL baselines, the ToMacVF achieves optimal performance and exhibits strong adaptability and robustness across various asynchronous multi-agent experimental scenarios.
Ablation studies further validate the contributions of each proposed module.
\end{itemize}

\setlength{\abovedisplayskip}{4pt}
\setlength{\belowdisplayskip}{4pt}

\begin{figure*}[ht]
\begin{center}
\centerline{\includegraphics[width=\linewidth]{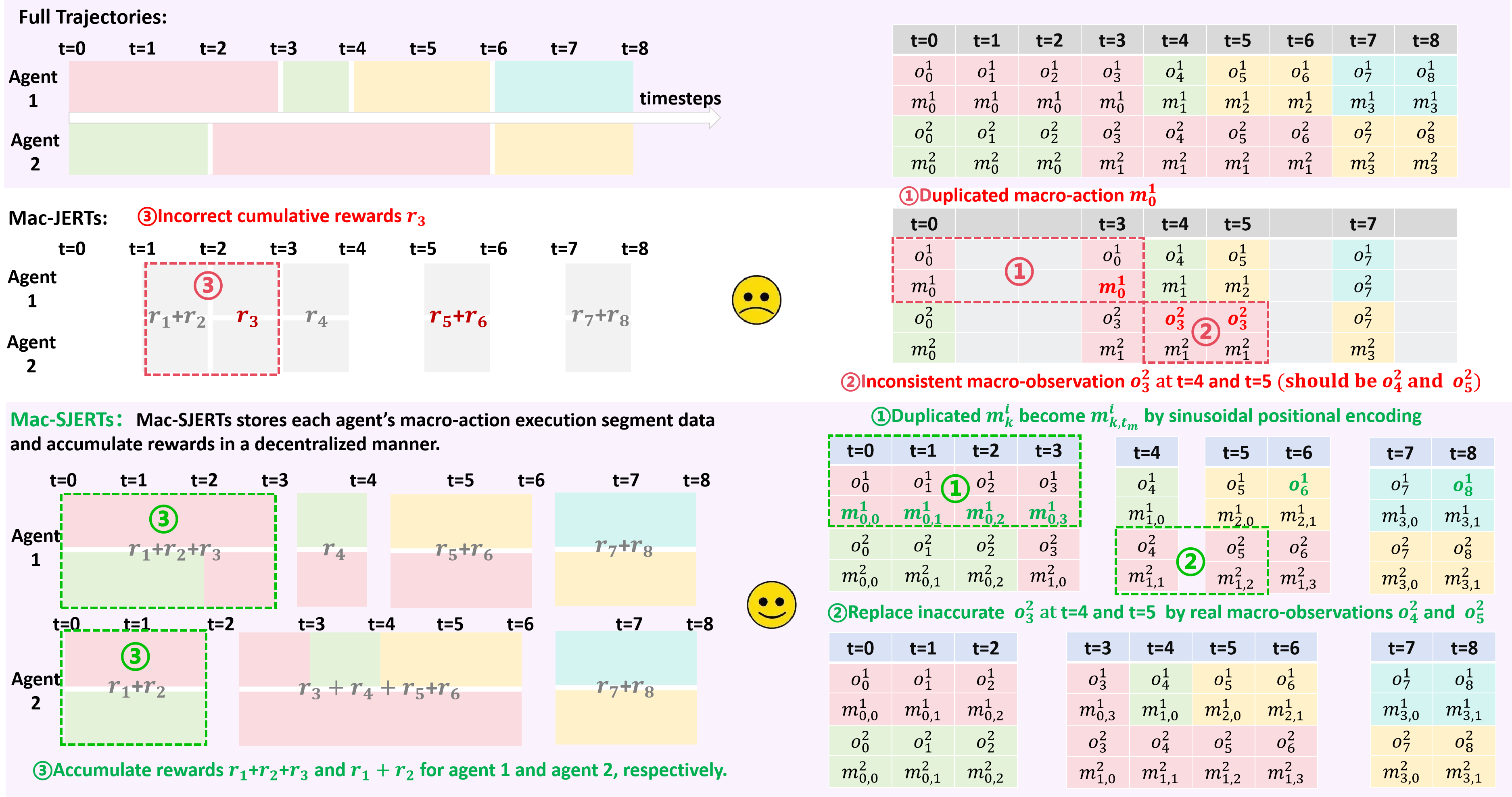}}
\caption{
Example of two training buffers in MacDec-POMDP.
To solve the issue of\textcolor{red}{\textcircled{1} duplicated macro-actions},
we utilize \textcolor[rgb]{0.0, 0.5, 0.0}{\textcircled{1} sinusoidal positional encoding to incorporate the temporal information into the original macro-actions}.
To address \textcolor{red}{\textcircled{2} the inconsistency between macro-observations in Mac-JERTs and the real ones},
we \textcolor[rgb]{0.0, 0.5, 0.0}{\textcircled{2}sample real macro-observations at each timesteps}. 
To eliminate \textcolor{red}{\textcircled{3} the incorrect cumulative rewards}(where the red cumulative rewards are incorrectly assigned, while the gray ones are correct), 
we construct \textcolor[rgb]{0.0, 0.5, 0.0}{\textcircled{3} independent buffer for each agent to store its own macro action execution segment data and accumulate rewards in a decentralized manner.} 
}

\label{rollout scheme}
\end{center}
\end{figure*}
\vspace{-0.3cm}
\section{Dec-POMDPs and MacDec-POMDPs}
\paragraph{Dec-POMDPs}
The decision-making problem of the cooperative multi-agent system can be formulated as a decentralized partially observable Markov decision process (Dec-POMDP) \cite{dec-pomdp}, 
which is denoted by a tuple ${\langle  \mathcal{I}, \mathcal{S}, \mathcal{U}, T_s, R, \mathcal{O},T_\mathcal{O}, \gamma  \rangle}$. 
$\mathcal{I}  =  \{1, 2, \cdots, n\}$ is a finite set consisting of $n$ agents.
$S$ is the state space.
$\mathcal{U} \equiv \langle U^i \rangle_{i \in \mathcal{I}}$ represents the joint action space, with $U^i$ denoting the action space of agent $i$.
$T_s:S\times\boldsymbol{\mathcal{U}}\times S\rightarrow[0,1] $ is the transition function, defined as $T_s(s_t,\boldsymbol{u}_t, s_{t+1}) = P(s_{t+1}|s_t,\boldsymbol{u}_t)$. 
All agents share the reward function $R: S   \times   \mathcal{U} \rightarrow   \mathbb{R}$. 
$\mathcal{O} \equiv \langle O^i \rangle_{i \in \mathcal{I}}$ represents the joint observation space, with $O^i$ denoting the observation space of agent $i$.
$T_\mathcal{O}: S \times \mathcal{U} \times \mathcal{O} \rightarrow[0,1]$ is the observation probability function, defined the probability of receiving a joint observation $\boldsymbol{o_{t+1}}$ when a joint action $\boldsymbol{u_t}$ were taken and arriving in state $s_{t+1}$, where $T_\mathcal{O} (\boldsymbol{o_{t+1}}, \boldsymbol{u_{t}},s_{t+1} ) = P  (\boldsymbol{o_{t+1}}| \boldsymbol{u_{t}},s_{t+1} )$.
$\gamma   \in   [0,1)$ is a discounted factor.

At time step $t$, agents receive an observation $\boldsymbol{o_t} \equiv \langle o^i_t \rangle_{i \in \mathcal{I}}  \in  \mathcal{O}$ according to the observation probability function $T_\mathcal{O}$ 
and each agent $i$ takes an action $u_t^i \in  U^i$ based on its policy $\pi^i(\cdot|s_t)$, 
where the global state $s_t$ is approximated by the history trajectories $h_t^i=\{o_1^i, u_1^i,\cdots, o_t^i, u_t^i\} \in   \mathcal{H}^i$.
This forms the joint action $\boldsymbol{u_t} \equiv \langle u^i_t \rangle_{i \in \mathcal{I}} \in  \mathcal{U}$
and the joint policy $\boldsymbol{\pi} \equiv \langle \pi^i \rangle_{i \in \mathcal{I}}$. 
Then, the next state $s_{t+1}$ is obtained through the transition function $T_s$. 
The shared reward $r_t$ is obtained through the reward function $R(s_t,\boldsymbol{u}_t)$. 
The joint policy $\boldsymbol{\pi}$ induces a normalized discounted state visitation distribution $d^{\boldsymbol{\pi}}$, where $d^{\boldsymbol{\pi}}(s) = (1-\gamma)\sum_{t=0}^{\infty}\gamma^tPr(s_t= s|\boldsymbol{\pi})$ and $Pr(\cdot |\boldsymbol{\pi} ):S \mapsto [0,1]$ is the state probability function under $\boldsymbol{\pi}$.
The optimization objective of joint policy is to maximize the expected cumulative return, denoted as $\boldsymbol{\pi}^* = \arg\max_{\boldsymbol{\pi}}\mathcal{J}(\boldsymbol{\pi})=\arg\max_{\boldsymbol{\pi}}\mathbb{E}_{s_0 \sim d^{\boldsymbol{\pi}}, (s_t,\boldsymbol{u}_t ) \sim (T_s, \boldsymbol{\pi}})[\sum_{t=0}^\infty\gamma^tR(s_t, \boldsymbol{u}_t)]$.

\paragraph{MacDec-POMDPs} Dec-POMDPs with temporally extended macro-actions are referred to as MacDec-POMDPs \cite{macPOMDP2}, which augments the DecPOMDPs tuple with 
$\langle\mathcal{M},\hat{\mathcal{O}},T_{\hat{o}^{i\in \mathcal{I}}},\hat{T_s}\rangle$:
where $\mathcal{M} \equiv \langle M^i \rangle_{i \in \mathcal{I}}$
and $\hat{\mathcal{O}} = \langle \hat{O}^i \rangle_{i \in \mathcal{I}}$ are the set of joint macro-actions and macro-observations.
Similar to the primitive case, we define joint macro-action-macro-observation trajectories
$\boldsymbol{\hat{h}} \in \hat{\boldsymbol{\mathcal{H}}}$
and local ones $\hat{h}^i \in \hat{\mathcal{H}}^i.$
Each agent’s $i$ macro-action $m^i$ is not the combination of micro-actions and is defined as a tuple
$\langle I_{m^i}, \pi_{m^i}, \beta_{m^i} \rangle$:
$I_{m^i} \subset \hat{\mathcal{H}}^i$ is the initiation set;
$\pi_{m^i} (\cdot | h^i_t)$ is the macro-action low-level policy, which is pre-trained via supervised learning using expert datasets;
$\beta_{m^i} : \mathcal{H}^i \rightarrow [0, 1]$ is the termination condition.
It is important to note that, in the asynchronous MARL training process, we train only macro-action policies directly for asynchronous decision-making, without training low-level policies $\pi_{m^i} (\cdot | h^i_t)$ for primitive micro-actions.
During decentralized execution, agents independently select macro-actions that form the joint one
$\boldsymbol{m} = \langle m^i \rangle_{i \in \mathcal{I}} \in \mathcal{M}$,
and maintain a high-level policy $\pi_{M^i} (m^i | \hat{h}^i)$. 
Taking into account the stochastic termination of a macro-action, the transition probability is rewritten as $\hat{T_s}(s,t_{\boldsymbol{m}},\boldsymbol{m}, s^{\prime}) = P(s^{\prime},t_{\boldsymbol{m}}|s,\boldsymbol{m})$, where $t_{\boldsymbol{m}}$ is the number of timesteps taken by the joint macro-action $\boldsymbol{m}$ that terminates when any agent completes its own macro-action;
Upon terminating its macro-action, the agent $i$ receives a macro-observation $\hat{o}^i \in \hat{O}^i$
according to a macro-observation probability function $T_{\hat{o}^i}: O^i \times M^i \times S \rightarrow [0, 1]$,
defined as $T_{\hat{o}^i} (\hat{o}^{i\prime}, m^i, s^{\prime}) \equiv P(\hat{o}^{i\prime} | m^i, s^{\prime})$. 
The overall objective is to find a joint high-level policy $\pi_{\mathcal{M}} (\boldsymbol{m} | \boldsymbol{\hat{h}})$ that maximizes the expected discounted return.

\paragraph{Value Factorization in Synchronous MARL}
In synchronous MARL, it is infeasible to train each agent separately or treat all agents as a single entity for joint training.
To address this challenge, the CTDE paradigm \citep{ctde} allows all agents to access global information in the centralized training and execute only based on local histories in a decentralized manner. 
Then, the Individual-Global Max(IGM) \citep{qplex} is proposed to guarantee that the optimality of individual policies is consistent with the optimality of joint policies, allowing greedy local actions to result in optimal global actions:
\begin{equation}
\arg\max_{a_t} Q_{total}(s_t, a_t) =
\{ \arg\max_{a^i_t} Q_i(h^i_t, u^i_t)|i \in \mathcal{I}
 \}
\end{equation}
% The objective of value decomposition methods \citep{wqmix,qtran,qplex} that follow IGM is to obtain an optimal joint action-value function based on the Bellman operator \citep{belleman}: $Q_{tot}(s_t,a_t)=r_t+\gamma\mathbb{E}_{s_{t+1}}[\max_{a_{t+1}}Q_{tot}(s_{t+1},a_{t+1})]$. 
% During centralized training, Q-functions $\{ Q_i |i \in \mathcal{I}\}$ are trained by minimizing the TD error:
During centralized training, the objective of value factorization methods \citep{wqmix,qtran,qplex} that follow IGM is to obtain an optimal joint action-value $Q^*_{total}$ by minimizing the TD error:
\begin{equation*}
\begin{split}
    \mathcal{L}_{TD}(\boldsymbol{\theta},\phi) = \mathbb{E}[r_t+\gamma\max\overline{Q}_{total}(s_{t+1},\boldsymbol{u}_{t+1})-Q_{tot}(s_t,\boldsymbol{u}_t)]^2 \\
     Q_{total}(s_t,\boldsymbol{u}_t)=\mathcal{F}_{mix}(Q_1(h_t^1,u^1_t;\theta^1),\cdots,Q_n(h_t^n,u^n_t;\theta^n),s_t;\phi)
     \end{split}
\end{equation*}
where $\boldsymbol{\theta}=\langle \theta^i\rangle_{i \in \mathcal{I}}$ are the parameters of the individual action-value networks $=\langle  Q_i\rangle_{i \in \mathcal{I}}$ , and $\phi$ is the parameters of the mixing network $\mathcal{F}_{mix}$.
$\overline{Q}_{total}$ is the target network for the joint action-value network, periodically copied from $Q_{tot}$. 
Previous value factorization methods in synchronous settings, such as QMIX \citep{wqmix}, Qtran \citep{qtran}, and QPLEX \citep{qplex}, make efforts to discover effective factorization structures for designing various mixing networks $\mathcal{F}_{mix}$.
\vspace{-0.1cm}
\section{Temporal Macro-Action Value Factorization (ToMacVF) Framework}
Efficient and fine-grained temporal asynchronous value factorization motivates a need for a principled way for end-to-end training of distributed asynchronous policies. 
In Section 3.1, Mac-SJERT is proposed to enable a comprehensive and precise representation of the macro-action execution process, providing essential information support for subsequent fine-grained value factorization.
To achieve principled and fine-grained asynchronous value factorization, the To-Mac-IGM is formalized in Section 3.2.
Based on the To-Mac-IGM, Section 3.3 presents the modularized architecture of the ToMacVF framework, which can conveniently integrate previous value factorization methods.
% 为了完整准确的宏动作过程表征和准确宏动作价值评估，3.1节提出了Mac-SJERT
% 为了实现principled and fine-garined asynchronous value factorization, 3.2节形式化了Temporal Macro-Action-Based IGM 
% 在3.3节，结合Mac-SJECT并基于To-Mac-IGM原则，我们提出了一种模块化的ToMacVF架构，并给出了一种结合qmix的具体实现。

\subsection{Macro-Action Segmented Joint Experience Replay Trajectory (Mac-SJERT)}
Previous asynchronous MARL methods aim to directly apply synchronous MARL approaches, excessively simplifying macro-actions into primitive actions.
This oversimplification is reflected in the design of the training buffers, Mac-JERTs, which adopt a centralized manner to collect joint trajectories and simply sample limited and biased information only at each macro-action termination timestep.

%缺点：Mac-JERTs冗余宏动作、不一致宏观测+累计折扣奖励和不准确
Specifically, Mac-JERTs encounter the problems of duplicated macro-actions, inconsistent macro-observations, and incorrect cumulative rewards.
%缺点：冗余宏动作表示
Duplicated macro-actions are caused by identical macro-actions with different execution progressions being represented in the same way.
We define $m_k^i$ to represent the $k$-th macro-action taken by agent $i$.
For instance, in Figure \ref{rollout scheme}\textcolor{red}{\textcircled{1}}, agent $1$'s macro-action $m_0^1$ has the same representation at $t=0$ and $t=3$.
% 举例说明有偏差的观测信息问题
Duplicated and inconsistent macro-observations are due to an agent’s macro-observation being updated only when its own macro-action terminates, leading to inconsistencies between macro-observations and the real ones at non-terminal timesteps.
For example, in Figure \ref{rollout scheme}\textcolor{red}{\textcircled{2}}, since agent $2$'s macro-action $m_1^2$ is still executing at $t=4$ and $t=5$, the macro-observation $o_3^2$ is not updated.
As a result, the macro-observation $o_3^2$ at $t=4$ and $t=5$ does not align with the real macro-observation $o_4^2$ and $o^2_5$.
% 举例说明累计折扣奖励不准确问题
Inaccurate cumulative rewards occur when the termination of any other agent’s macro-action resets the joint cumulative rewards to zero, leading to an underestimation of individual macro-action values.
As shown in Figure \ref{rollout scheme}\textcolor{red}{\textcircled{3}}, the joint cumulative reward is reset to zero due to the termination of agent 2's macro-action $m_0^2$ at $t=2$, preventing accurate reward assignment for agent 1's macro-action $m_0^1$. 
In fact, the correct joint cumulative reward should be $r_1+r_2+r_3$, but $m_0^1$ is only assigned $r_3$ when it terminates at $t=3$.

\begin{figure}[htbp]
\centerline{\includegraphics[width=0.9\linewidth]{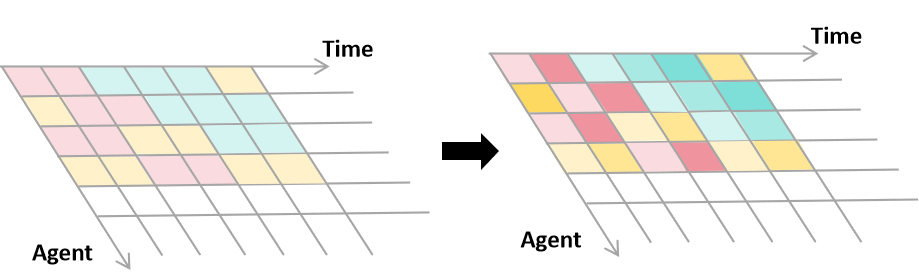}}
\caption{
Macro-action representation with temporal information. 
The horizontal axis represents time, while the vertical axis represents different agents. 
The cells are colored to distinguish between different macro-actions, and another feature introduced by temporal positional encoding is the shading intensity of each cell, which reflects the execution progress of the corresponding macro-action.
}
\label{macro-action}
\vspace{-7pt}
\end{figure}

To address these issues, a novel training buffer, termed Macro-Action Segmented Joint Experience Replay Trajectory (Mac-SJERT), is designed in Figure \ref{rollout scheme}.
The core idea of Mac-SJERT is to continuously update and collect the origin, real-time information at each timestep in a decentralized manner, in contrast to Mac-JERT, which stores outdated and inaccurate information in a centralized manner.
Mac-SJERT constructs individual buffers for each agent to store joint trajectory segments corresponding to complete macro-action executions and accumulate joint rewards in a decentralized manner (Figure \ref{rollout scheme}\textcolor[rgb]{0.0, 0.5, 0.0}{\textcircled{3}}).
The buffer for each agent $i$, denoted as $(\mathcal{D}^i, \prec)$, consists of a finite set of segmented joint experience replay trajectories for macro-actions. 
$\prec$ is the binary relation on $\mathcal{D}^i$, i.e. $\mathcal{D}^i_j\prec\mathcal{D}^i_k $ \emph{iff} $m^i_j$ is terminated before $m^i_k$ .
Each $ \mathcal{D}_k^i$ records the complete execution process of macro-action $m_k^i$ from start to end.
At each timestep $t$ of macro-action $m_k^i$, $ \mathcal{D}{_k^i} $ records the real information(Figure \ref{rollout scheme}\textcolor[rgb]{0.0, 0.5, 0.0}{\textcircled{2}}), including global state, joint macro-observations, joint macro-actions, and immediate joint rewards: $\left\{ s_{t}, \boldsymbol{\hat{o}}_{t}, \boldsymbol{m}_{t}, r_{t} \right\}$.
At the termination timestep of macro-action $m_k^i$, $ \mathcal{D}{_k^i} $ not only records the above-mentioned information but also additionally stores joint \textbf{cumulative} rewards accumulated over the entire duration of macro-action $m^i_k$: $\left\{ s_{t}, \boldsymbol{\hat{o}}_{t}, \boldsymbol{m}_{t}, r_{t}, R_{\boldsymbol{m}} \right\}$.
%这里特别的介绍宏动作位置编码
\paragraph{Temporal positional encoding}
To prevent duplicated macro-actions and macro-observations, we incorporate temporal information into them with sinusoidal positional encoding(Figure \ref{rollout scheme}\textcolor[rgb]{0.0, 0.5, 0.0}{\textcircled{1}}). 
The encoded time-step information $f_e(t)$ is concatenated with the macro-action $m^i_k$ and macro-observation $\hat{o}^i$, resulting in new encoded vectors:
\begin{eqnarray}
   {m^i_{k,t}} = m^i_k \oplus f_e(t); \ \ \ \ {\hat{o}^i_{t}} = \hat{o}_t^i \oplus f_e(t)
\end{eqnarray}
Additionally, to achieve a fine-grained representation of macro-actions as shown in Figure \ref{macro-action}, we distinguish macro-actions at different execution stages by encoding the execution progress timestep $t_{m^i_k}$ into their representation:
\begin{eqnarray}
  & {m^i_{k,t}} = m^i_k \oplus f_e(t) \oplus f_e(t_{m^i_k})
\end{eqnarray}
where $t_{m^i_k}$ represents the elapsed time from the start of the current macro-action to the current timestep, indicating the progress of the macro-action's execution.
% In the following, we use $m^i_{t}$ to replace $m_k^i$ to represent the macro action.

\paragraph{Micro-transitions and Macro-transitions} By utilizing the segmented joint experience replay trajectories in $\mathcal{D}_i$, we can construct two types of transitions as training datasets: micro-transitions for the timestep level and macro-transitions for the macro-action level. 
Micro-transitions with one timestep can be represented as:
\begin{equation}
    \mathcal{T}^{t:t+1} = \left\{s_{t}, \hat{\boldsymbol{o}_{t}}, \boldsymbol{m}_{t}, r_{t},s_{t+1},\hat{\boldsymbol{o}}_{t+1} \right\}
\end{equation}
where $\boldsymbol{m_t}$ and $\hat{\boldsymbol{o}}_t$ denote the joint macro-actions and macro-observations with temporal position encoding at timestep $t$, respectively.

Macro-transitions of macro-action $m_k^i$ can be represented as:
\begin{equation}
    \mathcal{T}_{m_k^i} = \left\{s, \hat{\boldsymbol{o}}, \boldsymbol{m},R_{\boldsymbol{m}},{s}^\prime, \hat{\boldsymbol{o}}^\prime \right\}
\end{equation}
where ${s}$, $\hat{\boldsymbol{o}}$, and $\boldsymbol{m}$  are the initial macro-state, joint macro-observations and joint macro-actions for $m_k^i$.
$R_{\boldsymbol{m}} =\sum_{t=0}^{t_{m^i_k}^{\text{max}}}\gamma^t r_t$ is the joint cumulative reward from start to end of $m_k^i$, where $t_{m^i_k}^{\text{max}}$ is the total duration of $m^i_k$. ${s}^\prime$ and $\hat{\boldsymbol{o}}^\prime$ are the final macro-state and joint macro-observations at the termination of $m_k^i$. All the joint macro-observations and joint macro-actions are represented with temporal positional encoding.

As a result, our Mac-SJERT ensures both the comprehensiveness and accuracy of macro-action execution information, while also providing more precise reward allocation across the entire macro-action duration.
It provides a complete and accurate representation of each macro-action execution process for subsequent fine-grained asynchronous value factorization.

\subsection{Temporal Macro-Action-Based IGM}
To achieve principled and fine-grained temporal asynchronous value factorization, it is essential to ensure consistent greedy selection between joint and individual macro-action value functions. 
Given that macro-actions often span multiple timesteps, it becomes necessary to define the space of both ongoing and terminated macro-actions.
For centralized training, we adopt the conditional prediction method from previous approaches \cite{asy_marl1}, applying the $\arg\max$ operator only to agents with terminated macro-actions that need to select new ones, while maintaining the ongoing macro-actions of other agents. 
Conversely, during decentralized execution, only agents with terminated macro-actions select new ones based on their local information.
Then, the Temporal Macro-Action-Based IGM (To-Mac-IGM) principle is formalized, incorporating the dimension of macro-action execution timesteps.
To-Mac-IGM provides a more generalized formulation compared to previous methods, improving the ability to capture the temporal dynamics throughout macro-action execution.
All proofs of theorems in this section are provided in the appendix.

\begin{theorem}[Defination of To-Mac-IGM ]
Given a joint macro-history $\hat{\boldsymbol{h}}=\{\hat{h}^i\}_{i=1}^{n} \in \hat{\boldsymbol{H}}$, we define the set of macro-action termination spaces $\mathcal{M}^+$ as:
\begin{eqnarray}
    \mathcal{M}^+=\{\mathcal{M}^i \in \mathcal{M}|\beta_{m^i \sim \pi_{\mathcal{M}^i}(\cdot|\hat{h}^i)} = 1, \, \forall i \in I\}
\end{eqnarray}
We define $T = \{0,\cdots,t_{max}\}$ is the execution timestep space of a macro-action, $t_{max}$ is maximum execution timesteps, $t_{m^i}$ is the execution time step of macro-action $m^i$, $\boldsymbol{t_m}=\{t_{m^i}|i\in\mathcal{I}\}$, and $\boldsymbol{m^{-}}$ are the macro-actions that are still executing.
For a joint macro-action-value function $Q_{total}:\hat{\boldsymbol{H}} \times \mathcal{M} \times T^n \rightarrow \mathbb{R}$, if there exist individual macro-action-value functions $[Q_i:\hat{H}^i \times M^i \times T \rightarrow \mathbb{R}]_{i=1}^n$, such that the following holds:
\begin{eqnarray}
    &\!\arg \max_{\boldsymbol{m \in \mathcal{M}}} Q_{total}(\hat{\boldsymbol{h}},\boldsymbol{m},\boldsymbol{t_m}|\boldsymbol{m}^-) =&\\ \nonumber
    &\!\begin{cases}\!
    \arg \!\max_{m^i \in \mathcal{M}^i} \!Q_i(\hat{h}^i,m^i,t_{m^i}\!=\!0),\! &\text{if } \mathcal{M}^i \!\in\! \mathcal{M}^+ \\
    m^i \in \boldsymbol{m^{-}}, &\! \text{otherwise}
    \end{cases} \!
    &\!\forall i \!\in\! I \nonumber
\end{eqnarray}
then $[Q_i(\hat{h}^i,m^i,t_{m^i})]_{i=1}^n$ satisfies \textbf{To-Mac-IGM} for $Q_{total}(\hat{\boldsymbol{h}},\boldsymbol{m},\boldsymbol{t_m}|\boldsymbol{m}^-)$ under $\hat{\boldsymbol{h}}$.
In this case, $Q_{total}$ is factorized by $[Q_i]_{i=1}^n$, and $[Q_i]_{i=1}^n$ are factors of $Q_{total}$.
\end{theorem}
\begin{figure*}[ht]
\centerline{\includegraphics[width=0.95\linewidth]{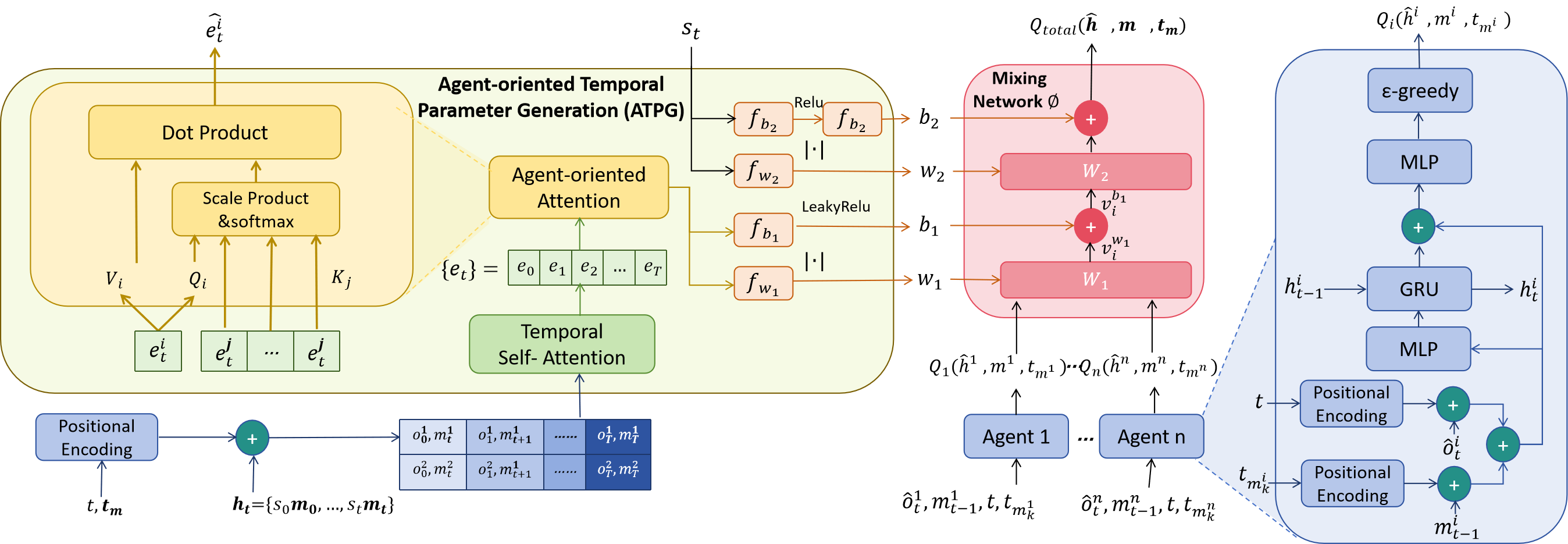}}
\vspace{-0.2cm}
\caption{
The network architecture of the ToMacVF.
}
\label{Framework}
\vspace{-0.2cm}
\end{figure*}
\vspace{-0.2cm}
Definition of To-Mac-IGM ensures that, for agents whose macro-actions have terminated, the greedy macro-action selection remains consistent between the centralized $Q_{total}$ and decentralized $Q_i$. 
Dec-POMDP can be viewed as a special case of MacDec-POMDP, where the primitive actions in IGM can be seen as the macro-actions in To-Mac-IGM with a single timestep.
Therefore, IGM is a simplified special case of To-Mac-IGM with macro-actions executed in a single timestep, and To-Mac-IGM can be generalized to IGM. 
As a result, To-Mac-IGM represents a more comprehensive class of functions than the original IGM and Mac-IGM:
\begin{theorem}
\label{theorem2}
    Denoting with 
    \begin{eqnarray}
        &\!F^\textit{IGM}\!=\!\{ \!( Q^\textit{IGM}\!:\!\boldsymbol{H} \!\!\times \!\mathcal{U} \!\!\rightarrow\!\! \mathbb{R}^{|\mathcal{U}|}\!,\!
        <\!Q_i^{IGM}\!:\!H^i\!\!\times \!U^i \!\!\rightarrow \!\!\mathbb{R}^{|U^i|\!}\!>_{i \in I} ) \!\} \nonumber\\
        &\!F^\textit{Mac-IGM}=\{ ( Q^\textit{Mac-IGM}:\hat{\boldsymbol{H}} \times \mathcal{M} \rightarrow \mathbb{R}^{|\mathcal{M}|}, \nonumber
        \\&\!<Q_i^\textit{Mac-IGM}:\hat{H}^i\times \mathcal{M}^i \rightarrow \mathbb{R}^{|\mathcal{M}^i|}>_{i \in I} ) \}\nonumber\\
        &\!F^\textit{To-Mac-IGM}=\{ ( Q^\textit{To-Mac-IGM}:\hat{\boldsymbol{H}} \times \mathcal{M} \times T^n \rightarrow \mathbb{R}^{|\mathcal{M}|\times|T^n|}, \nonumber\\
        &\!<Q_i^\textit{To-Mac-IGM}:\hat{H}^i\times \mathcal{M}^i \times T \rightarrow \mathbb{R}^{|\mathcal{M}^i|\times |T|}>_{i \in I} ) \}\nonumber
    \end{eqnarray}
    the class of functions satisfying IGM, Mac-IGM, and To-Mac-IGM, respectively, then:
    \begin{eqnarray}
    F^\textit{IGM} \subset F^\textit{Mac-IGM} \subset F^\textit{To-Mac-IGM}
    \end{eqnarray}
\end{theorem}
% To-MacAdv-IGM
The appendix also defines the equivalent transformation of To-Mac-IGM, named To-MacAdv-IGM, which transitions to temporal macro-action-based advantage functions.

\subsection{Implementation of ToMacVF Framework}
Previous AVF methods utilize the Mac-JERTs and directly apply previous synchronous value factorization methods to asynchronous settings without modification.
However, these methods fail to effectively capture the temporal dynamics and interdependence among agents during macro-action execution, leading to unsuitable and incorrect credit assignments to individual macro-action value functions in a coarse-grained manner.
For simplicity, we denote the macro-action $m^i_{k,t}$ at timestep $t$ as $m_t^i$ throughout this section.

To achieve more fine-grained and efficient value factorization, our ToMacVF framework incorporates Mac-JERTs and is designed based on the To-Mac-IGM.
The architecture of the ToMacVF Framework is illustrated in Figure \ref{Framework}.
It consists of an Agent-oriented Temporal Parameter Generation(ATPG) module, the temporal agent networks generating local utilities $Q_i$, and a mixing network.
In this architecture, temporal agent networks are designed to output fine-grained temporal individual macro-action value functions by utilizing temporal positional encoding(Section 3.1) to macro-observation $\hat{o}^i_t$ and macro-action $m^i_{t-1}$ at the last timestep $t-1$.
The ATPG module plays a critical role in satisfying the sufficiency of To-Mac-IGM by generating non-negative parameters for the mixing network.
By embedding temporal interdependence information into the policy gradient, the ATPG module ensures appropriate credit assignment and enhances effective inter-agent coordination.
Based on the parameters generated by the ATPG module, the mixing network calculates the joint macro-action value $Q_{total}$.
Notably, the mixing network can be conveniently implemented by previous value factorization methods, such as QMIX \cite{qmix}, Qtran \cite{qtran}, or QPLEX \cite{qplex}.
Finally, we implement the ToMacVF framework by integrating it with the QMIX mixing network.
Based on the CTDE, during decentralized execution, the mixing network and ATPG module are removed, and only the temporal agent networks are utilized to act greedily in a decentralized manner.
% It enables effective end-to-end training of asynchronous decentralized macro-action policies.
Pseudocode can be found in Appendix.
\paragraph{Agent-oriented temporal parameter generation(ATPG)}
The Agent-oriented temporal parameter generation(ATPG) module encodes historical trajectory segments using the temporal self-attention module and the agent-oriented self-attention module.
These encoded representations, along with the global state, are then used as inputs to generate the parameters for the mixing network.
By capturing the temporal dynamics and interactions among agents, this module provides effective support for the subsequent asynchronous value factorization.

During centralized training, the temporal self-attention module processes the historical trajectory $\{(s_t,\boldsymbol{m}_t)\}_{t=0}^T$ with causality masks to produce the temporal representation $\{e_t\}_{t=0}^T$, where $e_t=\{e_t^i\}_{i=0}^n$ represents the set of temporal encodings for each agent $i$.
Specifically, we have
\begin{eqnarray}
    &e_t^i=\text{Self\_Attn}(q_{i,t}^{TEM},k_{i}^{TEM},v_{i}^{TEM}) \nonumber\\
    =&\text{softmax}(\frac{q_{i,t}^{TEM}\cdot {k_{i}^{TEM}}^T}{\sqrt{d}} \odot M^{TEM})\cdot v_{i}^{TEM}
\end{eqnarray}
where $\odot$ is a Hadamard product\cite{hadamard}, the query matrix $q_{i,t}^{TEM}$ is obtained by applying the trainable matrix $W^{TEM}_Q$ to the output of an MLP that encodes macro-observations and macro-actions embeddings of agent $i$ at timestep $t$: $q_{i,t}^{TEM}=W^{TEM}_Q[\text{MLP}(\hat{o}_t^i,m_{k,t}^i)]$. 
Similarly, the key and value matrices are generated by the trainable matrices $W^{TEM}_K$ and $W^{TEM}_V$, respectively. 
The self-attention weight is calculated by applying a softmax function to the scaled dot product of the query and key matrices, with the scaling factor $\sqrt{d}$ and the causality mask $M^{TEM}$. 
$M^{TEM}$ is a causality mask with its first $t$-th entries equal to 1, and remaining entries 0. 
It ensures that at each timestep $t$, the module only attends to the states and macro-actions before $t$.
\begin{figure*}[ht]
\centering
\includegraphics[width=\linewidth]{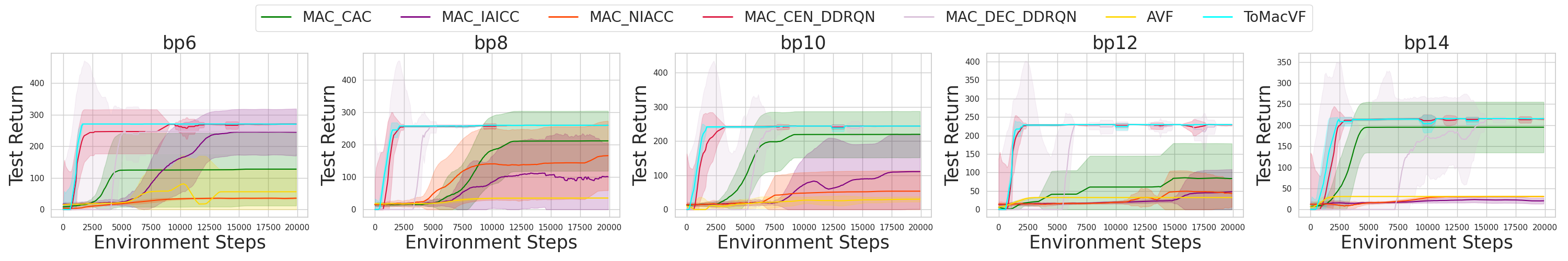}
\vspace{-0.6cm}
\caption{
The comparison of performance between ToMacVF and other baselines in BoxPushing.}
\label{fig:bp}
\end{figure*}

\begin{figure*}[ht]
\vspace{-0.5cm}
\centering
\includegraphics[width=\linewidth]{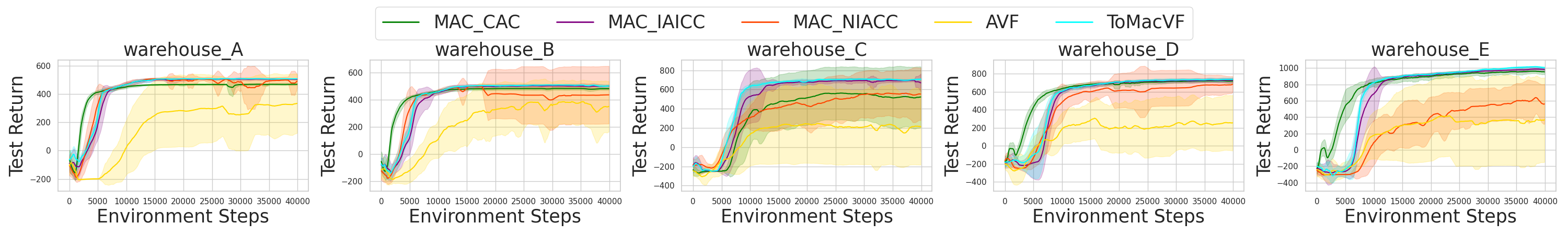}
\vspace{-0.6cm}
\caption{The comparison of performance between ToMacVF and other baselines in Warehouse.}
\label{fig:wh}
\vspace{-0.3cm}
\end{figure*}
Next, the agent-oriented self-attention module is applied to capture the interdependence among agents.
The query vector for agent $i$ at timestep $t$ is denoted as $q_{i,t}=W_Q\cdot e_t^i$, where $e_t^i$ is generated by the temporal self-attention module, and $W_Q$ is a trainable query projection matrix. 
Similarly, the key and value vectors of agent $i$ at timestep $t$ are denoted as $k_{i,t}=W_K\cdot e_t^i$, $v_{i,t}=W_V\cdot e_t^i$, respectively.
The attention weight $\alpha_t^{ij}$, which captures the correlation between agents $i$ and $j$ at timestep $t$, is computed as:
\begin{eqnarray}
    \alpha_t^{ij}=\text{softmax}(\frac{q_{i,t}\cdot {k_{j,t}}^T}{\sqrt{d}})
\end{eqnarray}
where $\alpha_t^{ii}=0$. In addition, if $\alpha_t^{ij}<\delta $, $\alpha_t^{ij}=0$, where $ \delta >0$ is a hyper-parameter indicated the minimum degree of interdependence. 
The final output encodings $\hat{\boldsymbol{e}_t}=\{\hat{e}_t^i\}$ of the module can be computed as follows:
\begin{eqnarray}
    \hat{e}_t^i = \text{softmax}(\frac{q_{i,t}\cdot \boldsymbol{{k_{t}}^T}}{\sqrt{d}})\cdot \boldsymbol{v_{t}}
\end{eqnarray}

The final encodings $\hat{\boldsymbol{e}_t}=\{\hat{e}_t^i\}$  are used as inputs to the parameter generators, which produce the non-negative weights $\mathrm{w}_1$ and biases $b_1$ of the mixing network.
The non-negativity of these parameters ensures the monotonicity of value factorization, thereby satisfying the sufficiency condition of To-Mac-IGM.
The parameter generator $f_{\mathrm{w}_1}$ embeds $\hat{\boldsymbol{e_t}}$ into the policy gradient as follows:
\begin{eqnarray}
    \frac{\partial Q_{total}}{\partial Q_i} = \frac{\partial Q_{total}}{\partial v_i^{\mathrm{w}_1}} \cdot \frac{\partial v_i^{\mathrm{w}_1}}{\partial Q_i} =  \frac{\partial Q_{total}}{\partial v_i^{\mathrm{w}_1}} \cdot f_{\mathrm{w}_1}(\hat{\boldsymbol{e}}_t)
\end{eqnarray}
where $v_i^{\mathrm{w}_1}$ is the representations after $\mathrm{w}_1$ marked in Figure \ref{Framework}.
ATPG module provides temporal interdependence information for $Q_{total}$ generation that facilitates efficient inter-agent coordination.

\paragraph{Overall training}
Unlike previous asynchronous MARL methods, which use a coarse-grained approach to evaluate and update policies at macro-action termination, we introduce fine-grained value factorization and update policies at each timestep.
Based on the two transitions constructed by Mac-SJERT, we design two TD loss functions: micro-TD loss and macro-TD loss.
For each training step, the ToMacVF algorithm minimizes the micro-TD loss:
\begin{eqnarray}
    &\!\mathcal{L}_{micro-TD}(\boldsymbol{\theta},\phi)=\mathbb{E}_{\mathcal{D}=\langle D_i\rangle_{i=0}^n}[(r_t+\\
    &\!\gamma\max \bar{Q}_{total}(\hat{\boldsymbol{h}}_{t+1},\boldsymbol{m_{t+1}},\boldsymbol{t_{m_{t+1}}}) \nonumber \!-\!Q_{total}(\hat{\boldsymbol{h}}_{t},\boldsymbol{m_t},\boldsymbol{t_{m_t}}))^2]
\end{eqnarray}
After each agent $i$ terminates its macro-action $m^i_k$, the ToMacVF minimizes the following Macro-TD error: 
\begin{eqnarray}
    &\mathcal{L}_{Macro-TD}(\boldsymbol{\theta},\phi)=\mathbb{E}_{\mathcal{D}_i}[(R_{\boldsymbol{m}}+\\
    &\gamma^{t^{\text{max}}_{m^{i}_k}} \max\bar{Q}_{total}(\hat{\boldsymbol{h}}',\boldsymbol{m}',\boldsymbol{t_{m'}}) \nonumber 
    -Q_{total}(\hat{\boldsymbol{h}},\boldsymbol{m},\boldsymbol{t_m}))^2]
\end{eqnarray}
where  $\bar{Q}_{total}$ is a target network with periodic updates.
The micro-TD loss promotes effective collaboration at the fine-grained level of macro-actions, while the macro-TD loss enhances collaboration at higher dimensions. 
By updating the TD loss at both levels, the algorithm enables more precise and comprehensive credit assignment over the agent's macro-action decision-making process.

\section{Experiment}
The experiments aim to address the following questions:
i) Is ToMacVF more effective than existing asynchronous baselines in various asynchronous collaboration scenarios?
ii) Can the ATPG module assist in effective asynchronous value factorization?
iii) Does our Mac-SJERT buffer provide an effective representation of the macro-action execution process for asynchronous value factorization?
iiii) How does integrating ToMacVF with different value factorization methods affect its performance?

To follow prior works and enable comparison, we conducted experiments in three open-source standard experimental environments for MacDec-POMDP: BoxPushing \cite{boxpushing}, Warehouse Tool Delivery (WTD) \cite{warehouse2}, and Overcooked \cite{overcooked}. 
We conduct a comparison of the ToMacVF algorithm with several classic asynchronous baselines, including Mac-CAC\cite{asy_marl3}, Mac-IAICC, Mac-NIACC \cite{macros}, Mac-Dec-MADDRQN, Mac-Cen-DDRQN \cite{warehouse2}, and AVF \cite{avf}. 
However, given the poor performance of Mac-Dec-MADDRQN and Mac-Cen-DDRQN in the Warehouse scenarios reported in previous studies \cite{macros}, their results are excluded from this paper.
Considering the importance of statistically significant results \cite{statistical_comparisons}, we report the smoothed average return over 10 episodes of 20 independent runs for each method. 
Detailed descriptions of each experimental scenario and setting are provided in the Appendix.
\begin{figure}[ht]
\centering
\includegraphics[width=\linewidth]{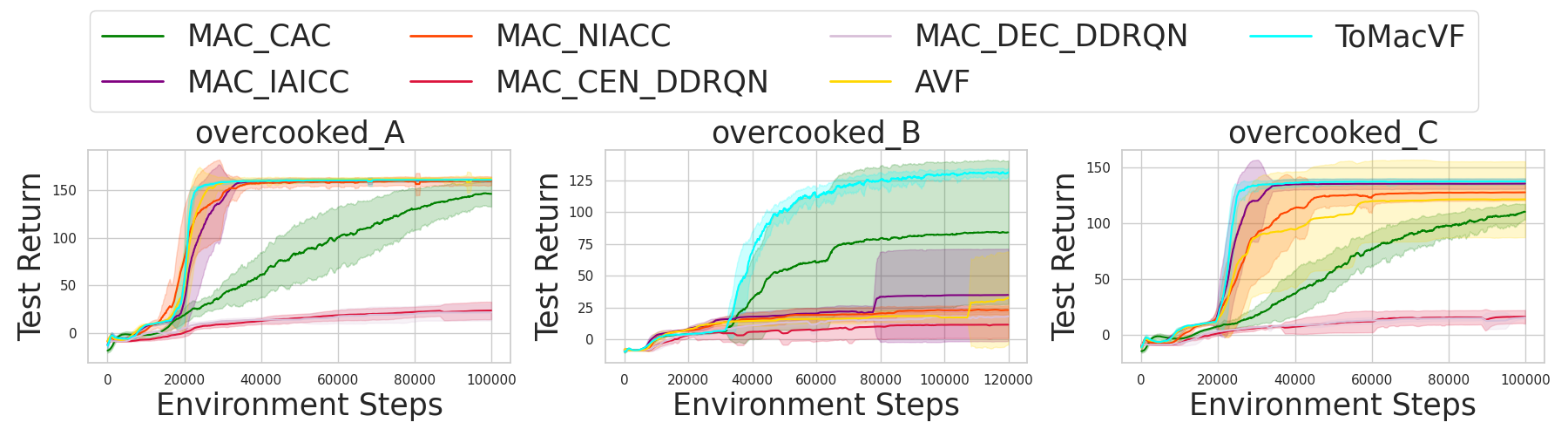}
\vspace{-0.6cm}
\caption{The comparison of performance between ToMacVF and other baselines in Overcooked.}
\label{fig:ovc}
\end{figure}
\paragraph{ToMacVF algorithm performance} 
The experimental results are presented in Figures \ref{fig:bp}-\ref{fig:ovc}.
Although Q-learning-based methods such as Mac-Dec-MADDRQN and Mac-Cen-DDRQN achieve strong performance in the BoxPushing scenarios, prior research has shown that these approaches exhibit poor adaptability in more complex asynchronous environments like Overcooked and Warehouse \cite{macros}, where they obtain the lowest returns.
In contrast, Mac-IAICC and Mac-NIACC deliver comparable performance in the Warehouse and Overcooked scenarios but fail to generalize well to larger BoxPushing maps, where their performance significantly degrades.
AVF performs poorly with high variance in Warehouse and Overcooked, and even collapses to zero return in the BoxPushing scenarios, due to its incorrect and coarse-grained credit assignments in a coarse-grained manner.
In comparison, our proposed ToMacVF algorithm demonstrates strong adaptability across diverse tasks. It consistently delivers high performance across all scenarios, showing superior stability, faster convergence, and robust generalization to different asynchronous coordination challenges.

\begin{figure}[ht]
\centering
\includegraphics[width=\linewidth]{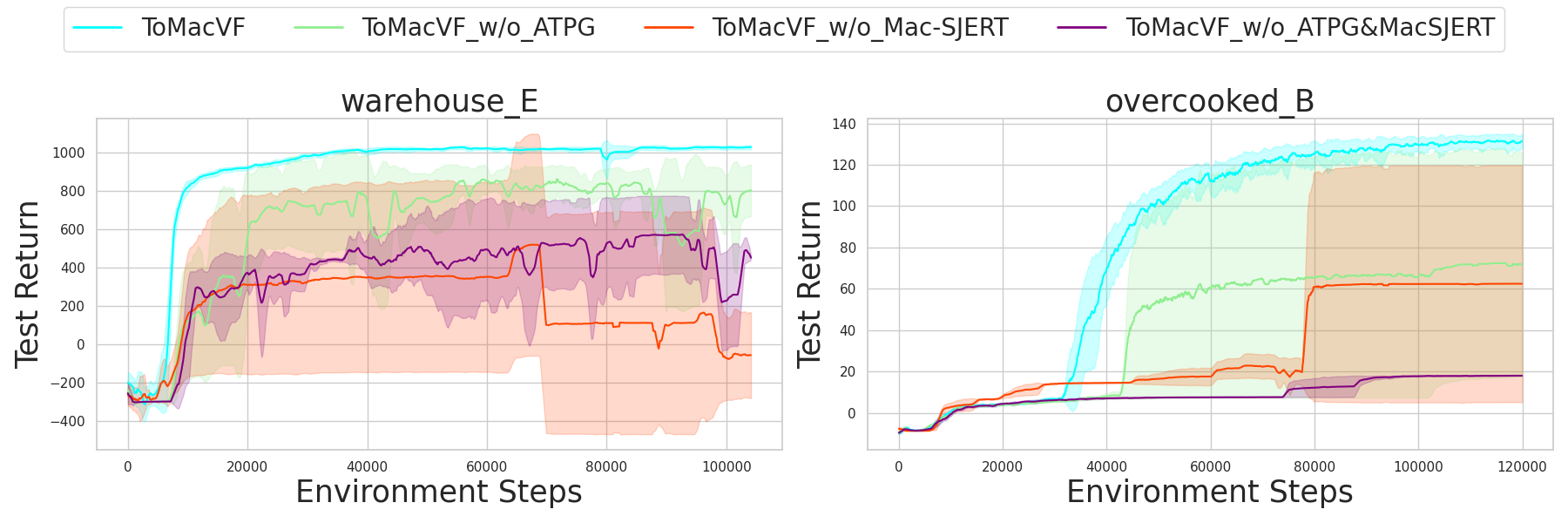}
\vspace{-0.4cm}
\caption{Ablations on Mac-SJERT and ATPG modules on two challenging maps.}
\label{fig:ablations}
\vspace{-0.2cm}
\end{figure}

\paragraph{Ablation studies on Mac-SJERT and ATPG}
To further validate the contribution of each core component, we conduct ablation studies on the Mac-SJERT buffer and the ATPG module in two of the most challenging asynchronous scenarios: overcooked\_B and warehouse\_E.
As illustrated in Figure \ref{fig:ablations}, ToMacVF\_w/o\_ATPG denotes the variant where the ATPG module is removed, and the original QMIX architecture is used without parameter generation from ATPG.
ToMacVF\_w/o\_Mac\_SJERT replaces our proposed Mac-SJERT buffer with the conventional Mac-JERTs, thus discarding the richer intermediate information during macro-action execution.
ToMacVF\_w/o\_ATPG\&Mac\_SJERT combines both modifications, simultaneously removing the ATPG module and using Mac-JERTs in place of Mac-SJERT.
The results in Figure \ref{fig:ablations} clearly show that both the ATPG module and Mac-SJERT buffer play critical roles in boosting learning efficiency and final performance. 
Specifically, the ATPG module captures fine-grained temporal interdependencies across asynchronous agent behaviors, enhancing collaboration among agents effectively under partial observability. 
Meanwhile, the Mac-SJERT buffer enables more complete and accurate representations of the macro-action execution process, which are essential for effective asynchronous value factorization. 
Together, these components allow the agents to learn more efficient and cooperative policies in asynchronous complex environments.
\begin{figure}[ht]
\centering
\includegraphics[width=\linewidth]{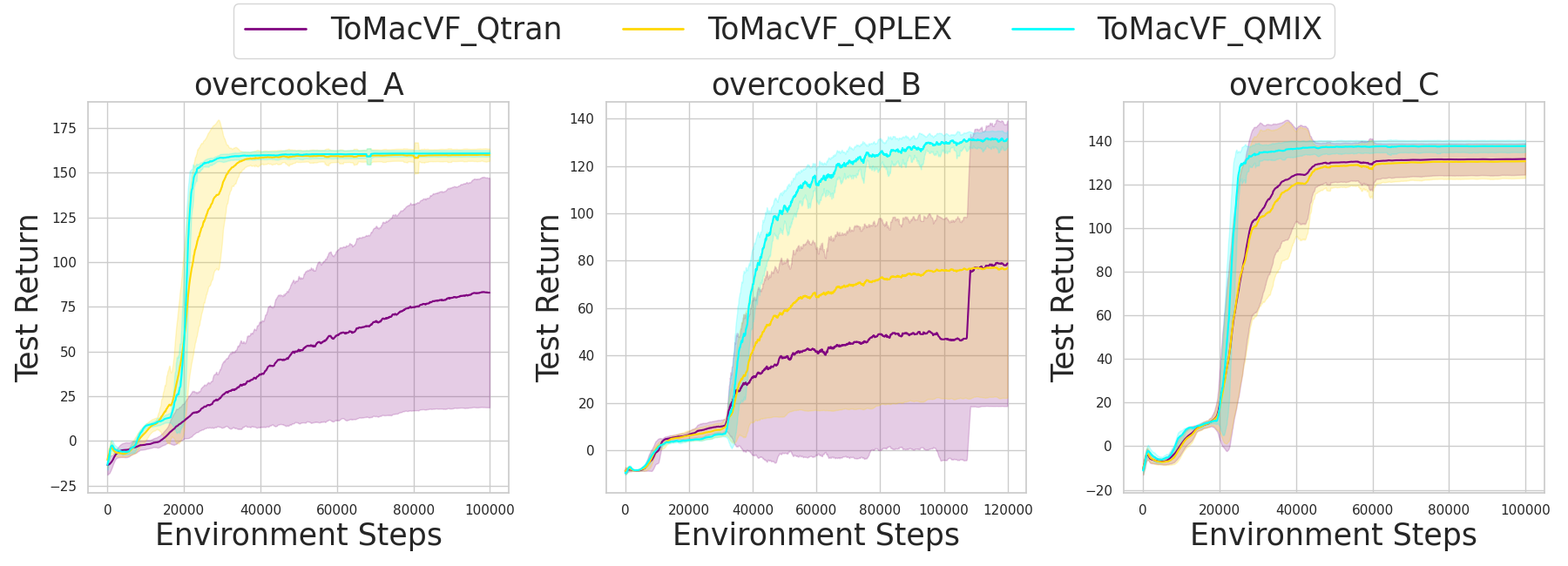}
\vspace{-0.5cm}
\caption{Ablations on integrating ToMacVF with various value factorization methods.}
\label{fig:ablations_VF}
\end{figure}
\paragraph{Ablation studies on integrating ToMacVF with various value factorization methods}
We also investigate the impact of integrating ToMacVF with different value factorization methods in the Overcooked environment.
As shown in Figure \ref{fig:ablations_VF}, ToMacVF integrated with QMIX (ToMacVF\_QMIX) consistently achieves strong performance across all scenarios, demonstrating robust generalization. 
In contrast, ToMacVF integrated with Qtran (ToMacVF\_Qtran) and with QPLEX (ToMacVF\_QPLEX) performs poorly in some scenarios and suffers from high variance, indicating unstable training and limited adaptability under asynchronous settings.
These results highlight the importance of choosing a value factorization method that aligns well with our ToMacVF in asynchronous multi-agent coordination tasks.
\section{Related Works}
Value factorization methods in MARL have made remarkable achievements but struggle in asynchronous scenarios, where agents execute macro-actions of varying durations and initiate or terminate these actions at different timesteps.
Classic asynchronous MARL methods, including Mac-CAC \cite{asy_marl3}, Mac-IAICC, Mac-NIACC \cite{macros}, Mac-Dec-MADDRQN, and Mac-Cen-DDRQN \cite{warehouse2}, have been proposed to address this challenge. 
These methods typically leverage centralized training, utilizing buffers such as macro-action joint experience replay trajectories (Mac-JERTs) \cite{warehouse1}. 
Mac-JERTs treat macro-actions as primitive actions by sampling limited and often biased information only at macro-action termination timesteps, leading to incomplete and inaccurate value estimations due to the loss of temporal details in the execution process. 
While these methods train macro-action policies with Mac-JERTs and conventional synchronous MARL techniques, they overlook crucial temporal details inherent in macro-action execution, leading to poor performance due to incomplete and inaccurate value estimations.
Recent asynchronous value factorization (AVF) methods \cite{avf} also adopt Mac-JERT and directly apply synchronous techniques in asynchronous environments, leading to incorrect and coarse-grained credit assignments.
In contrast, our method overcomes these limitations by designing a framework specifically for asynchronous settings, preserving fine-grained temporal value factorization based on a more comprehensive and precise representation of the macro-action process.
\section{Conclusion}
In this work, we focus on addressing challenges in asynchronous Multi-Agent Reinforcement Learning.
First, a centralized training buffer, Mac-SJERT, is introduced to enable more comprehensive and precise representations of the macro-action execution process, providing effective data for subsequent asynchronous value factorization.
Additionally, the consistency between joint and individual macro-action greedy selection is formalized as To-Mac-IGM, demonstrating that it can be generalized to synchronous cases.
The ToMacVF framework is designed based on the To-Mac-IGM and equipped with Mac-SJERT to enable fine-grained temporal credit assignment, facilitating the effective end-to-end training of asynchronous decentralized macro-action policies.
Crucially, the proposed ToMacVF framework can be extended by any value factorization structure.
Overall, experimental results demonstrate that the ToMacVF algorithm successfully achieves optimal performance and exhibits strong adaptability and robustness across various asynchronous multi-agent experimental scenarios.

\newpage
\appendix
\onecolumn
\section{Appendix}
\subsection{Pseudocode}
\begin{algorithm}[ht]
\caption{Temporal Macro-action Value Factorization}
\begin{algorithmic}[1]
\State \textbf{Input:}
\State \hspace{1em} Decentralized agent and target agent networks $\langle Q_{\theta_i}, Q_{\theta_i'} \rangle_{i \in \mathcal{N}}$
\State \hspace{1em} Mixer and target mixer networks $Q_\phi, Q_{\phi'}$
\State \hspace{1em} Decentralized memory buffers $\mathcal{D}=\{\mathcal{D}_i\}$, initialized with Mac-SJERTs
\State \hspace{1em} Initial macro observations and macro-states $\langle o^i_t, s^i_t \rangle_{i \in \mathcal{N}}$
\State \hspace{1em} Number of episodes $e$, time limit per episode $t'$, update frequency $f$, and target network update coefficient $\omega$
\State \hspace{1em} Centralized $Q_\Theta$ and target $Q_{\Theta'}$ networks for training: $\langle \langle Q_{\theta_i} \rangle_{i \in \mathcal{N}}, Q_\phi \rangle$ and $\langle \langle Q_{\theta_i'} \rangle_{i \in \mathcal{N}}, Q_\phi' \rangle$
\For {episode = $1$ to $e$}
 \For {timestep $t = 0$ to $t'$}
\For {each agent $i \in \mathcal{N}$}
 \If {macro-action $m^i_t$ is terminated}
\State Select a new macro-action $m^i$ using $\epsilon$-greedy policy based on $Q_{\theta_i}(\hat{h}^i, m^i, t_{m^i} = 0)$
 \EndIf
\EndFor
\State Execute joint macro-action $\boldsymbol{m} = \{m^i\}_{i \in \mathcal{N}}$ in the environment
\State Receive joint reward $r_t$ at time step $t$
\For {each agent $i \in \mathcal{N}$}
\State Store transition $\mathcal{T}^{t:t+1}$ into $\mathcal{D}_i$ as in Section 3.1
\If {macro-action $m^i$ is terminated}
 \State Store transition $\mathcal{T}_{m^i}$ into $\mathcal{D}_i$ as in Section 3.1
 \State Reset joint cumulative macro-action reward $R_{\boldsymbol{m}}$ and execution time $t_{m^i}$
\EndIf

\EndFor
\If {$t \mod f = 0$}
 \For {each agent $i \in \mathcal{N}$}
\State Sample trajectories from $\mathcal{D}_i$
\State Compute per-agent action-values using $\langle Q_{\theta_i}, \theta_i, \hat{h}^i_t, m^i_t, t_{m^i} \rangle$
\State Use states $s_t$ and histories $\hat{h}_t$ to generate hyperparameters for mixer networks $Q_\phi, Q_{\phi'}$
\State Factorize action-values using mixer networks $Q_\phi, Q_{\phi'}$
\State Perform a gradient descent step on micro-TD $\mathcal{L}_{Micro-TD}(\Theta)$ following Eq. 12 
\If {macro-action $m^i_t$ is terminated}
 \State Perform an additional gradient descent step on macro-TD loss $\mathcal{L}_{Macro-TD}(\Theta)$ following Eq. 13
\EndIf
\State Update target network weights: $\Theta' \gets \omega \Theta' + (1 - \omega) \Theta$
 \EndFor
\EndIf
 \EndFor
\EndFor
\end{algorithmic}
\end{algorithm}
\subsection{Defination of To-Mac-IGM and To-MacAdv-IGM}
\paragraph{Theorem 1}(Definition of To-Mac-IGM).
Given a joint macro-histories $\hat{\boldsymbol{h}}=\{\hat{h}^i\}_{i=1}^{n} \in \hat{\boldsymbol{H}}$, we define the set of macro-action termination spaces $\mathcal{M}^+$ as:
\begin{eqnarray}
 \mathcal{M}^+=\{\mathcal{M}^i \in \mathcal{M}|\beta_{m^i \sim \pi_{\mathcal{M}^i}(\cdot|\hat{h}^i)} = 1, \, \forall i \in I\}
\end{eqnarray}
We define $T = \{0,\cdots,t_{max}\}$ is the execution timestep space of a macro-action, $t_{max}$ is maximum execution timesteps, $t_{m_i}$ is the execution time step of macro-action $m_i$.
For a joint macro-action-value function $Q_{total}:\hat{\boldsymbol{H}} \times \mathcal{M} \times T^n \rightarrow \mathbb{R}$, if there exist individual macro-action-value functions $[Q_i:\hat{H}^i \times M^i \times T \rightarrow \mathbb{R}]_{i=1}^n$, such that the following holds:
\begin{eqnarray}
 &\!\arg \max_{\boldsymbol{m \in \mathcal{M}}} Q_{total}(\hat{\boldsymbol{h}},\boldsymbol{m},\boldsymbol{t_m}|\boldsymbol{m}^-) =&\\ \nonumber
 &\!\begin{cases}\!
 \arg \!\max_{m^i \in \mathcal{M}^i} \!Q_i(\hat{h}^i,m^i,t_{m_i}\!=\!0),\! &\text{if } \mathcal{M}^i \!\in\! \mathcal{M}^+ \\
 m^i \in \boldsymbol{m^{-}}, &\! \text{otherwise}
 \end{cases} \!
 &\!\forall i \!\in\! I \nonumber
\end{eqnarray}
then $[Q_i(\hat{h}^i,m^i,t_{m_i})]_{i=1}^n$ satisfies \textbf{To-Mac-IGM} for $Q_{total}(\hat{\boldsymbol{h}},\boldsymbol{m},\boldsymbol{t_m}|\boldsymbol{m}^-)$ under $\hat{\boldsymbol{h}}$.
In this case, $Q_{total}$ is factorized by $[Q_i]_{i=1}^n$, and $[Q_i]_{i=1}^n$ are factors of $Q_{total}$.

\paragraph{Theorem 2}(Definition of To-MacAdv-IGM).
Given joint macro-histories $\hat{\boldsymbol{h}}=\{\hat{h}^i\}_{i=1}^{n} \in \hat{\boldsymbol{H}}$ and $\mathcal{M}^+$, 
for a joint macro-action-value function $Q_{total}:\hat{\boldsymbol{H}} \times \mathcal{M} \times T^n \rightarrow \mathbb{R}$, 
defines as $Q_{total}(\hat{\boldsymbol{h}},\boldsymbol{m},\boldsymbol{t_m})=V(\hat{\boldsymbol{h}})+A(\hat{\boldsymbol{h}},\boldsymbol{m},\boldsymbol{t_m})$ ,
if there exist individual macro-action-value functions $[Q_i:\hat{H}^i \times \hat{M}^i \times T \rightarrow \mathbb{R}]_{i=1}^n$, 
defines as $Q_i(\hat{h}^i,m^i,t_{m_i})=V_i(\hat{h}^i)+A_i(\hat{h}^i,m^i,t_{m_i})$ , such that the following holds:
\begin{eqnarray}
 &\arg \max_{\boldsymbol{m \in \mathcal{M}}} A_{total}(\hat{\boldsymbol{h}},\boldsymbol{m},\boldsymbol{t_m}|\boldsymbol{m}^-) =&\\ \nonumber
 &\begin{cases}\!
 \arg \!\max_{m^i \in \mathcal{M}^i} A_i(\hat{h}^i,m^i,t_{m_i}\!=\!0) , & \!\text{if } \mathcal{M}^i \!\in\! \mathcal{M}^+ \\
 m^i \in \boldsymbol{m^{-}}, & \!\text{otherwise}
 \end{cases} \!
 &\! \forall i \!\in\! I \nonumber
\end{eqnarray}
then we say $[Q_i(\hat{h}^i,m^i,t_{m_i})]_{i=1}^n$ satisfies \textbf{To-MacAdv-IGM} for $Q_{total}(\hat{\boldsymbol{h}},\boldsymbol{m},\boldsymbol{t_m}|\boldsymbol{m}^-)$ under $\hat{\boldsymbol{h}}$.

\subsection{Representational Complexity of To-Mac-IGM and To-MacAdv-IGM}
Conventional value decomposition methods, such as VDN and QMIX\cite{qmix}, have limitations in their ability to represent functions accurately, making it difficult to fully capture the true value function category. 
Moreover, in partially observable environments (e.g., Dec-POMDP and MacDec-POMDP), the partial observations can lead to inconsistencies between individual action selections and the joint optimal action, especially in macro-action scenarios,  which are prone to value sorting errors:
\begin{equation}
 Q_i(\hat{h^i},m^i) > Q_i(\hat{h^i},m^{i \prime}) \ \text{when} \ Q(s,(\boldsymbol{m}^{-i},m^i))<Q(s,(\boldsymbol{m}^{-i},m^{i\prime}))
\end{equation}

where $ \boldsymbol{m}^{-i}$ is the joint action of all the agents excluding agent $i$.
To solve this problem, we combine state information with temporal interdependence information into the mixing network during centralized training, thereby improving the accuracy of the individual value functions. 
To overcome the challenge of insufficient historical information, we introduce recurrent neural networks (RNNs) in individual networks and integrate attention mechanisms with RNNs in the mixing network, enabling the storage of additional temporal historical data. 
This enhancement strengthens both temporal modeling capabilities and the effectiveness of value decomposition.

\subsection{Representational Expressiveness of To-Mac-IGM and To-MacAdv-IGM Algorithms}
%% The file named.bst is a bibliography style file for BibTeX 0.99c
The proposed To-Mac-IGM and To-MacAdv-IGM methods improve existing value factorization methods (such as VDN, QMIX, and QPLEX) by incorporating temporal information into the individual utility networks and mixing networks. 
We show that To-Mac-IGM is capable of encompassing a broader function space compared to IGM and Mac-IGM, while To-MacAdv-IGM also extends its representational capacity beyond that of Adv-IGM and MacAdv-IGM.
\paragraph{Theorem 3} Denoting with 
\begin{eqnarray}
 &F^\textit{IGM}=\{ ( Q^\textit{IGM}:\boldsymbol{H} \times \mathcal{U} \rightarrow \mathbb{R}^{|\mathcal{U}|}  \nonumber,
 <Q_i^{IGM}:H^i\times U^i \rightarrow \mathbb{R}^{|U^i|}>_{i \in I} ) \} \nonumber\\
 &F^\textit{Mac-IGM}=\{ ( Q^\textit{Mac-IGM}:\hat{\boldsymbol{H}} \times \mathcal{M} \rightarrow \mathbb{R}^{|\mathcal{M}|}, \nonumber
 <Q_i^\textit{Mac-IGM}:\hat{H}^i\times \mathcal{M}^i \rightarrow \mathbb{R}^{|\mathcal{M}^i|}>_{i \in I} ) \}\nonumber\\
 &F^\textit{To-Mac-IGM}=\{ ( Q^\textit{To-Mac-IGM}:\hat{\boldsymbol{H}} \times \mathcal{M} \times T^n \rightarrow \mathbb{R}^{|\mathcal{M}|\times|T^n|}, \nonumber
 <Q_i^\textit{To-Mac-IGM}:\hat{H}^i\times \mathcal{M}^i \times T \rightarrow \mathbb{R}^{|\mathcal{M}^i|\times |T|}>_{i \in I} ) \}\nonumber
\end{eqnarray}
the class of functions satisfying IGM, Mac-IGM, and To-Mac-IGM respectively, then:
\begin{eqnarray}
F^\textit{IGM} \subset F^\textit{Mac-IGM} \subset F^\textit{To-Mac-IGM}
\end{eqnarray}

\proof MacDec-POMDPs extend Dec-POMDPs by replacing the primitive actions available to each agent with option-based macro-actions. 
However, as shown in \cite{macPOMDP2}, the macro-action set must include primitive actions to guarantee the same globally optimal policy. Specifically, we have:
\begin{eqnarray}
U^i \subset M^i , \forall i \in I
\end{eqnarray}
Meaning that $\forall i \in I$, $|M^i| > |U^i|$, which implies $|\mathcal{M}| > |\mathcal{U}|$. 
It also follows that $\mathcal{O} \subseteq \hat{\mathcal{O}}$ as a MacDec-POMDP is, in the limit where only primitive actions are selected, equivalent to a DecPOMDP. 
Given these relationships, we can conclude that:$| \hat{\boldsymbol{H}} \times \mathcal{M}| > |\boldsymbol{H} \times \mathcal{U}|$ (i.e., the domain over which primitive action value functions are defined is smaller than the domain over which macro action value functions are defined). Hence, 
\begin{eqnarray}
F^\textit{IGM} \subset F^\textit{Mac-IGM}.
\end{eqnarray}
Similarly, for the domain of macro-action value functions with an temporal dimension, we have:  $| \hat{\boldsymbol{H}} \times \mathcal{M} \times T^n| > |\hat{\boldsymbol{H}} \times \mathcal{M}|$,where $T = \{0,\cdots,t_{{max}}\}$ is the execution timestep space of a macro-action, $t_{{max}}$ is maximum execution timestep.
This leads to the conclusion: 
\begin{eqnarray}
 F^\textit{Mac-IGM} \subset F^\textit{To-Mac-IGM}
\end{eqnarray}
Finally, combining the above results, we conclude that:
\begin{eqnarray}
F^\textit{IGM} \subset F^\textit{Mac-IGM} \subset F^\textit{To-Mac-IGM}
\end{eqnarray}

\paragraph{Theorem 4} The consistency requirement of To-MacAdv-IGM is equivalent to the To-Mac-IGM one. Hence, denoting with
\begin{eqnarray}
  F^\textit{To-MacAdv-IGM}= 
  \{ ( Q^\textit{To-MacAdv-IGM}:\hat{\boldsymbol{H}} \times \mathcal{M} \times T^n \rightarrow \mathbb{R}^{|\mathcal{M}|\times|T^n|},
 <Q_i^\textit{To-MacAdv-IGM}:\hat{H}^i\times \mathcal{M}^i \times T \rightarrow \mathbb{R}^{|\mathcal{M}^i|\times |T|}>_{i \in I} ) \}
\end{eqnarray}
the class of functions satisfying To-MacAdv-IGM, we can conclude that $F^{\textit{To-Mac-IGM}} \equiv F^\textit{To-MacAdv-IGM}$.
\proof Given a joint macro-history $\hat{\boldsymbol{h}} \in \hat{\boldsymbol{H}}$ on which $[Q_i(\hat{h}^i,m^i,t_{m_i})]_{i=1}^n$ satisfies To-Mac-IGM for $Q_{total}(\hat{\boldsymbol{h}},\boldsymbol{m},\boldsymbol{t})$, we show theorem 2 represents the same consistency constraint as theorem 1. 
By applying the dueling decomposition from \cite{wang2016dueling}, we know $Q_{total}(\hat{\boldsymbol{h}},\boldsymbol{m},\boldsymbol{t_m})=V(\hat{\boldsymbol{h}})+A(\hat{\boldsymbol{h}},\boldsymbol{m},\boldsymbol{t_m})$, and $Q_i(\hat{h}^i,m^i,t_{m_i})=V_i(\hat{h}^i)+A_i(\hat{h}^i,m^i,t_{m_i}), \ \forall i \in I$. Hence, the state-value functions defined over macro-histories do not influence the action selection process. For the joint value, we can thus conclude that:

\begin{eqnarray}
 \arg \max_{\boldsymbol{m} \in \mathcal{M}} Q(\hat{\boldsymbol{h}}, \boldsymbol{m} ,\boldsymbol{t_m}\mid \boldsymbol{m}^-) 
= \arg \max_{\boldsymbol{m} \in \mathcal{M}} V(\hat{\boldsymbol{h}}) + A(\hat{\boldsymbol{h}}, \boldsymbol{m} ,\boldsymbol{t_m} \mid \boldsymbol{m}^-) 
= \arg \max_{\boldsymbol{m} \in \mathcal{M}} A(\hat{\boldsymbol{h}}, \boldsymbol{m} ,\boldsymbol{t_m}\mid \boldsymbol{m}^-) 
\end{eqnarray}
Similarly, for the individual values:
\begin{eqnarray}
\forall i \in \mathcal{N}, &\quad& 
\begin{cases}
\arg \max_{m^i \in \mathcal{M}^i} Q_i(\hat{h}^i, m^i,t_{m_i}) & \text{if } M^i \in \mathcal{M}^+ \\ \nonumber
m^i \in m^- & \text{otherwise}
\end{cases} \\
&=& 
\begin{cases}
\arg \max_{m^i \in \mathcal{M}^i} V(\hat{h}^i) + A_i(\hat{h}^i, m^i,t_{m_i}) & \text{if } M^i \in \mathcal{M}^+ \\
m^i \in m^- & \text{otherwise}
\end{cases} \\
&= &
\begin{cases}
\arg \max_{m^i \in \mathcal{M}^i} A_i(\hat{h}^i, m^i,t_{m_i}) & \text{if } M^i \in \mathcal{M}^+ \\
m^i \in m^- & \text{otherwise} \nonumber
\end{cases}
\end{eqnarray}
Broadly speaking, we know the history values act as a constant for both the joint and local estimation and do not influence the arg max operator. By combining Eq. 11 and 12, we conclude the equivalence between To-MacAdv-IGM and To-Mac-IGM.

\paragraph{Theorem 5} Denoting with $F^{\{\textit{Adv-IGM}, \textit{MacAdv-IGM}, \textit{To-MacAdv-IGM}\}}$ the classes of functions satisfying Adv-IGM, MacAdv-IGM and To-MacAdv-IGM, respectively, then:
\begin{eqnarray}
  F^{\textit{IGM}} \equiv F^\textit{Adv-IGM} \subseteq F^{\textit{Mac-IGM}} \equiv F^\textit{MacAdv-IGM}
  \subseteq F^{\textit{To-Mac-IGM}} \equiv F^\textit{To-MacAdv-IGM}
\end{eqnarray}
\proof The result naturally follows from Theorem 3,4, and the result of \cite{qplex} that established the equivalence between the function classes represented by the primitive IGM and Adv-IGM.
In more detail, from the latter we know:
\begin{eqnarray}
 F^{\textit{IGM}} \equiv F^\textit{Adv-IGM}
\end{eqnarray}
The previous work\cite{avf} demonstrated the equivalence between the function classes represented by Mac-IGM and MacAdv-IGM:  
\begin{eqnarray}
 F^\textit{Mac-IGM} \equiv F^\textit{MacAdv-IGM}
\end{eqnarray}
Moreover, Theorem 3 showed us that 
\begin{eqnarray}
 F^\textit{IGM} \subset F^\textit{Mac-IGM} \subset F^\textit{To-Mac-IGM}
\end{eqnarray}
Additionally, Theorem 4 demonstrated the equivalence:  
\begin{eqnarray}
 F^{\textit{To-Mac-IGM}} \equiv F^\textit{To-MacAdv-IGM}\\
\end{eqnarray}
Combining these results,  we arrive at the conclusion presented in Eq. 13.
\subsection{Experiment Description}
\subsection{BoxPushing}

In this cooperative task, two agents are required to coordinate their actions to push a large box to a designated goal area located at the top of a grid world.
Successfully moving the large box to the goal yields a higher reward compared to the alternative task of individually pushing smaller boxes. 
While a single agent can move a small box independently, the large box necessitates the simultaneous effort of both agents.
The environment's state space encompasses the positions and orientations of the agents, as well as the locations of the boxes. 
Agents can execute a set of primitive actions, including moving forward, turning left or right, or remaining stationary.
Beyond these basic actions, macro-actions are also available, such as \textbf{Go-to-Small-Box(i)} and \textbf{Go-to-Big-Box}. 
These macro-actions navigate an agent to predefined waypoints (indicated by red markers) beneath the corresponding boxes, ensuring that the agent is correctly positioned and oriented for subsequent operations. 
The \textbf{Push} macro-action allows an agent to push forward and terminates if the agent hits a boundary or collides with the large box.

Agent observations are significantly constrained, with each agent only able to perceive the state of the grid cell directly in front of it. 
The cell can contain one of the following: an empty space, a teammate, a boundary, a small box, or a large box.
The team is rewarded with +100 for successfully pushing the large box to the goal area or +10 for delivering a small box to the goal area. 
Conversely, a penalty of -10 is incurred if an agent attempts to push the large box alone or if an agent collides with a boundary. 
An episode ends when any box reaches the goal area or when the maximum horizon of 100 timesteps is reached.

In this work, we explore a variation of this task by altering the size of the grid world, specifically considering grid sizes of 6×6, 8×8, 10×10, 12×12, and 14×14. 
These variations introduce additional challenges that increase the complexity of coordination and navigation required for the agents to successfully complete the task. 
As the grid world expands, agents must adapt to increased spatial reasoning demands and greater potential for miscoordination, thereby testing the robustness and scalability of their cooperative policies.

\subsection{Warehouse}
The Warehouse Tool Delivery environment involves a team of robots assisting a human in assembling an item through four sequential work phases. 
Each phase requires a specific tool, and the human starts with the tool for the first phase. 
For subsequent phases, the team of robots must locate and deliver the correct tools in the proper order to ensure the human can continue working without delays. 
The workspace is a 5 × 7 continuous area, consisting of a workshop where the human works and a tool room where tools are stored.

The environment includes mobile robots and a manipulator robot. 
The mobile robots navigate between the workshop and the tool room to deliver tools, while the manipulator robot is responsible for searching for the required tools on the table in the tool room and passing them to the mobile robots.
The manipulator robot places tools in a staging area on the table, which can hold up to two tools at a time. 
Each mobile robot has three macro-actions: \textbf{Go-W}, which navigates the robot to the workshop; 
\textbf{Go-TR}, which directs it to the tool room; 
and \textbf{Get-Tool}, which positions the robot beside the table to receive a tool. 
The manipulator robot has four macro-actions: \textbf{Wait-M}, which lasts one time step to wait for mobile robots; 
\textbf{Search-Tool(i)}, which takes six timesteps to locate a specific tool and place it in the staging area; 
\textbf{Pass-to-M(i)}, which costs four timesteps to pass a tool to a specific mobile robot; 
and \textbf{Idle}, which pauses the manipulator when the staging area is full.

The state space includes the 2D positions of mobile robots, the remaining time for the manipulator robot's current macro-action, the human’s current work phase and its completion progress, and the location of each tool. 
Observations for mobile robots are limited to their own location, the tool they are carrying, and the number of tools at waiting spots in the tool room or the staging area. 
In the workshop, a mobile robot can observe the current work phase of the human. 
The manipulator robot can only observe the tools in the staging area and the identity of any mobile robots waiting nearby.

Rewards are designed to encourage timely and accurate collaboration.
The team receives +100 for delivering the correct tool to the human in time, while a penalty of -20 is imposed for delayed deliveries. 
A penalty of -10 is given if the manipulator robot attempts to pass a tool to a mobile robot that is not adjacent, and a step penalty of -1 applies to discourage unnecessary delays. 
Each episode ends after 150 timesteps or when the human obtains the tool for the final work phase. 
This environment emphasizes efficient coordination and decision-making under partial observations to minimize delays and maximize productivity.

We consider five variations of WTD :a) WTD-D, two working humans with a slower first work phase and two mobile robots. b) WTD-D, two working humans with different speeds and two mobile robots. c) WTD-T, three working humans with slower work phases and two mobile robots; d) WTD-T1, three working humans with different speeds and three mobile robots; e) WTD-Q, four working humans with slower work phases and three mobile robots.
The human working speeds under different scenarios are listed in Table \ref{WTD}.
\begin{table}[h!]

\centering
\caption{The number of time steps each human takes on each working phase in scenarios.}

\begin{tabular}{llllll} 
\toprule
\textbf{Scenarios} & \textbf{WTD-D} & \textbf{WTD-D1} & \textbf{WTD-T} & \textbf{WTD-T1} & \textbf{WTD-Q} \\ 
\midrule
\textbf{Human-0} & {[}27, 20, 20, 20{]}  & {[}18, 15, 15, 15{]}  & {[}40, 40, 40, 40{]}  & {[}38, 38, 38, 38{]}  & {[}40, 40, 40, 40{]}\\ 
\textbf{Human-1} & {[}27, 20, 20, 20{]}& {[}48, 18, 15, 15{]}  & {[}40, 40, 40, 40{]}  & {[}38, 38, 38, 38{]} & {[}40, 40, 40, 40{]}\\ 
\textbf{Human-2}& N/A& N/A& {[}40, 40, 40, 40{]}  & {[}27, 27, 27, 27{]}  & {[}40, 40, 40, 40{]}\\ 
\textbf{Human-3}& N/A& N/A  & N/A  & N/A& {[}40, 40, 40, 40{]}\\
\bottomrule
\end{tabular}
\label{WTD}
\end{table}

\subsection{Overcooked}
\begin{figure}
 \centering
  \vspace{-0.5cm}
 \includegraphics[width=0.5\linewidth]{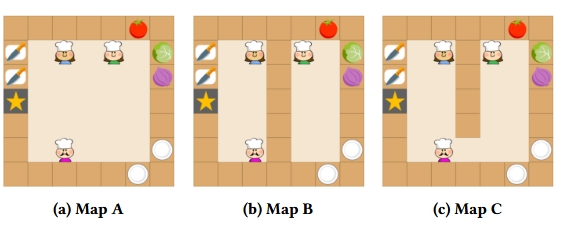}
 \caption{The collection of Overcooked Environments.}
 \label{fig:enter-label}
\end{figure}
The Overcooked environment, adapted from GymCooking, involves three agents collaboratively preparing and delivering salads as quickly as possible. 
To complete the task, agents must chop vegetables (such as tomatoes and onions), place the chopped ingredients on a plate, and deliver the finished salad to a designated destination, marked by a yellow star. 
The environment consists of three distinct 7×7 grid maps (A, B, and C), each offering unique challenges and layouts.

Each agent operates within a 5×5 observation grid centered around their position. They execute macro-actions, such as moving to specific ingredients, chopping, or delivering the salad. 
A macro-action is completed when the goal is achieved, or if the goal becomes unreachable. 
For example, in map A, the macro-action "go to tomato" ends when the agent reaches the tomato. 
In contrast, map B divides the environment into sections, where an agent in one section cannot reach ingredients in the other, causing the macro-action to terminate immediately.

The reward system encourages efficiency and accuracy.
Agents receive small rewards (+10) for chopping vegetables and a significant reward (+200) for delivering a correct salad (lettuce, tomato, and onion).
Penalties are applied for mistakes, such as delivering incorrect salads (-5), and minor penalties (-0.1) are given for each timestep spent during the task.

The episode ends either when agents successfully deliver the correct salad or when the time limit of 200 steps is reached. 
This environment challenges the agents to optimize their actions, work together effectively, and navigate a dynamic space, making it a valuable test for multi-agent reinforcement learning.

\subsection{Hyper-Parameters}

In this section, we list the hyper-parameters used for generating the results in this work. We choose the best performance of each method depending on its final converged value as the first priority.

\begin{table}[h!]

\centering

\caption{Hyper-parameters used for ToMacVF in Box Pushing $6 \times 6 - 14 \times 14$.}

\label{tab:hyperparameters}
\begin{tabular}{ccccccc}
\toprule
\textbf{Parameter}& \textbf{$6\times6$}& \textbf{$8\times8$}& \textbf{$10\times10$}& \textbf{$12\times12$}& \textbf{$14\times14$}\\ \midrule
Training Episodes& 40K & 40K & 40K & 40K & 40K\\ 
Learning rate & 0.001  & 0.001  & 0.001  & 0.0005 & 0.005\\ 
Batch size  & 128  & 128 & 128 & 128 & 128  \\ 
Train freq (episode)& 10  & 10& 14 & 16  & 16\\ 
Episodes per train& 16  & 16& 32& 48  & 48\\ 
Target-net update freq (episode) & 32& 32& 48& 144  & 144\\ 
rnn-layer size & 32& 32& 48& 144  & 144\\
$\epsilon_{\text{start}}$& 1& 1 & 1 & 1& 1 \\ 
$\epsilon_{\text{end}}$  & 0.1 & 0.1 & 0.1 & 0.1 & 0.1  \\ 
$\epsilon_{\text{decay}}$ (episode) & 4K & 4K& 4K& 4K  & 4K\\ \bottomrule
\end{tabular}

\end{table}

\begin{table}[h!]
\centering

\caption{Hyper-parameters used for ToMacVF in Overcooked map $A-C$.}

\label{tab:hyperparameters}
\begin{tabular}{ccccccc}
\toprule
\textbf{Parameter}& \textbf{map $ A$}& \textbf{map $ B$}& \textbf{map $ C$}\\ \midrule
Training Episodes& 100K & 120K & 100K \\ 
Learning rate & 0.0003  & 0.0003  & 0.0003 \\ 
Batch size  & 64  & 128 & 128  \\ 
Train freq (episode)& 32  & 32 & 64 \\ 
Target-net update freq (episode) & 16& 16& 16\\
rnn-layer size & 32& 32 & 32\\
N-step TD & 5  & 5& 5\\ 
$\epsilon_{\text{start}}$& 1& 1 & 1  \\ 
$\epsilon_{\text{end}}$  & 0.05 & 0.05 & 0.05   \\ 
$\epsilon_{\text{decay}}$ (episode) & 20K & 20K& 20K\\ \bottomrule
\end{tabular}

\end{table}

\begin{table}[h!]
\centering

\caption{Hyper-parameters used for ToMacVF in Warehouse $A-E$.}

\label{tab:hyperparameters}
\begin{tabular}{ccccccc}
\toprule
\textbf{Parameter}& \textbf{Warehouse $A$}& \textbf{Warehouse $B$}& \textbf{Warehouse $C$}& \textbf{Warehouse $D$}& \textbf{Warehouse $E$}\\ \midrule
Training Episodes& 40K & 40K & 80K & 80K & 100K\\ 
Learning rate & 0.0005  & 0.0005  & 0.0003 & 0.0003 & 0.0001\\ 
Batch size  & 32  & 32 & 32 & 32 & 32  \\ 
Train freq (episode)& 8 & 8& 8& 8  & 8\\ 
Target-net update freq (episode) & 32& 32& 32& 64  & 64\\ 
rnn-layer size & 64& 64& 64& 64  & 64\\
N-step TD & 5  & 5& 5& 5 & 5\\ 
$\epsilon_{\text{start}}$& 1& 1 & 1 & 1& 1 \\ 
$\epsilon_{\text{end}}$  & 0.05 & 0.05 & 0.01 & 0.01 & 0.01  \\ 
$\epsilon_{\text{decay}}$ (episode) & 10K & 10K& 10K& 10K  & 10K\\ \bottomrule
\end{tabular}

\end{table}

\begin{table}[h!]
\caption{The notations and symbols used in this paper.}
\vskip 0.01in
\centering
\begin{small}
\begin{tabular}{ll}
\toprule
Notation & Definition \\
\midrule
$S$ & The state space\\
$I$ & The finite set of agents\\
$n$ & The number of agents\\
$i$ & The agent index\\
$\boldsymbol{\mathcal{U}}$ & The action space of agents, where $\boldsymbol{\mathcal{U}} \equiv \langle U^i \rangle_{i \in \mathcal{I}}$\\
$U^i$ & The action space of agent $i$\\
$T_s$& The transition function,defined as $T_s(s_t,\boldsymbol{u}_t, s_{t+1}) = P(s_{t+1}|s_t,\boldsymbol{u})$\\
$R$& The shared reward function\\
$O^i$ & The observation space of agent $i$ \\
$\mathcal{O}$  & The observation space, where $\mathcal{O} \equiv \langle O^i \rangle_{i \in \mathcal{I}}$\\
$T_\mathcal{O}$  & The observation probability function, defined the probability of receiving a joint observation $\boldsymbol{o_{t+1}}$ when a joint action $\boldsymbol{u_t}$ \\
&were taken and arriving in state $s_{t+1}$, where $T_\mathcal{O} (\boldsymbol{o_{t+1}}, \boldsymbol{u_t},s_{t+1} ) = P  (\boldsymbol{o_{t+1}}| \boldsymbol{u_t},s_{t+1} )$\\
$\gamma$& The discount factor\\

$t$& The time-step $t$\\
$s_t$& The state at time-step $t$\\
$s_t^i$& The state of agent $i$ at time-step $t$\\
$o_t^i$  & The observation of agent $i$ at time-step $t$\\
$\boldsymbol{o_t}$ & The joint observation at time-step $t$, where $\boldsymbol{o_t} \equiv \langle o^i_t \rangle_{i \in \mathcal{I}}  \in  \mathcal{O}$\\
$u_t^i$& The action of agent $i$ at time-step $t$\\
$\boldsymbol{u}_t$& The joint action at time-step $t$, where $\boldsymbol{u_t} \equiv \langle u^i_t \rangle_{i \in \mathcal{I}} \in  \mathcal{U}$\\
$h_t^i$& The trajectory of agent $i$ at time-step $t$\\
$\mathcal{H}^i$ & The trajectory space of agent $i$, where $\mathcal{H}^i\equiv (O^i \times U^i)^{*}$\\
$\pi^i$& The policy of agent $i$\\
$\boldsymbol{\pi}$& The joint policy of agents, where $\boldsymbol{\pi} \equiv \langle \pi^i \rangle_{i \in \mathcal{I}}$\\
$r_t$ & The shared rewards of agents at time-step $t$, where $r_t=R(s_t,\boldsymbol{u_t})$\\
$Pr(\cdot|\pi)$& The state probability function under $\pi$\\
$d^{\boldsymbol{\pi}}$& The discounted state visitation distribution, where $d^\pi(s) = (1-\gamma)\sum_{t=0}^{\infty}\gamma^tPr(s_t= s|\pi)$\\

$\mathcal{M}$ & The space of joint macro-actions, where $\mathcal{M} \equiv \langle M^i \rangle_{i \in \mathcal{I}}$\\
$M^i$ & The macro-action space of agent $i$\\
$\hat{\mathcal{O}}$ & The space of joint macro-observations, where $\hat{\mathcal{O}} = \langle \hat{O}^i \rangle_{i \in \mathcal{I}}$\\
$\hat{O}^i$ & The macro-observation space of agent $i$\\
$\hat{o}^i$ & The macro-observation agent $i$, where$\hat{o}^i \in \hat{O}^i$\\
$\hat{\boldsymbol{o}}$ & The joint macro-observation of agents \\
$\hat{\boldsymbol{o}'}$ & The next joint macro-observation of agents \\
$\hat{h}^i$& The macro-action-macro-observation trajectory of agent $i$\\
$\hat{\mathcal{H}}^i$ & The macro-action-macro-observation trajectory space of agent $i$\\
$\boldsymbol{\hat{h}}$& The joint macro-action-macro-observation trajectories\\
$\hat{\boldsymbol{\mathcal{H}}}$ & The joint macro-action-macro-observation trajectory space of agents\\

$m^i$ & The macro-action of agent $i$ \\
$I_{m^i}$ & The initiation set of macro-action $m^i$\\
$\pi_{m^i}$ & The low-level policy of macro-action $m^i$\\
$\beta_{m^i}$ & The termination condition of macro-action $m^i$, where $\beta_{m^i} : \mathcal{H}^i \rightarrow [0, 1]$\\
$\boldsymbol{m}$ & The joint macro-action, where $\boldsymbol{m} = \langle m^i \rangle_{i \in \mathcal{I}}$\\
$\pi_{M^i}$ & The high-level policy of agent $i$ for making macro-action decisions\\
$\hat{T_s}$ & The transition probability rewritten as $\hat{T_s}(s,t_{\boldsymbol{m}},\boldsymbol{m}, s^{\prime}) = P(s^{\prime},t_{\boldsymbol{m}}|s,\boldsymbol{m})$, where $t_{\boldsymbol{m}}$ is the number of timesteps \\
&taken by the joint macro-action $\boldsymbol{m}$ that terminates when any agent completes its own macro-action \\
$T_{\hat{o}^i}$& The macro-observation probability function, where $T_{\hat{o}^i} (\hat{o}^{i\prime}, m^i, s^{\prime}) \equiv P(\hat{o}^{i\prime} | m^i, s^{\prime}) : O^i \times M^i \times S \rightarrow [0, 1]$\\
$\pi_{\mathcal{M}} $ & The joint high-level policy for making macro-action decisions\\

\bottomrule
\end{tabular}
\end{small}
\end{table}

\begin{table}[bp]
\caption{The notations and symbols used in this paper.}
\vskip 0.01in
\centering
\begin{small}
\begin{tabular}{ll}
\toprule
Notation & Definition \\
\midrule

$\mathcal{D}_i$ &  The finite set of segmented joint experience replay trajectories for agent $i$\\
$m^i_k$ & The $k$-th macro-actions executed by agent\\
$m^i_{k,t}$ & The $k$-th macro-actions executed by agent at timestep $t$ with temporal positional encoding\\
$ \mathcal{D}_k^i $ &  The joint experience segments of macro-action $m_k^i$\\
$R_{\boldsymbol{m}}$ & The joint cumulative reward from
start to end of the macro-action.\\
$m^i_t$ & The macro-action of agent $i$ with temporal positional encoding\\
$\hat{o}^i_t$ & The macro-observation of agent $i$ with temporal positional encoding\\
$t_{m^i}$ & The execution time step of macro-action $m^i$\\
$t_{m^i_k}$ & The execution time step of macro-action $m^i_k$\\
$\mathcal{T}_{t:t+1}$ &  The transition from timestep $t$ to $t+1$ of macro-action, where $\mathcal{T}^{t:t+1} = \left\{s_{t}, \boldsymbol{o}_{t}, \boldsymbol{m}_{t}, r_{t},s_{t+1},o_{t+1} \right\}$\\
$\mathcal{T}_{m_k^i}$ & The transition of macro-action $m_k^i$ from start to end, where $\mathcal{T}_{m_k^i} = \left\{\hat{s}, \hat{\boldsymbol{o}}, \boldsymbol{m},R_{\boldsymbol{m}},\hat{s}^\prime, \hat{\boldsymbol{o}}^\prime \right\}$\\

$\mathcal{M}^+$ & The macro-action termination spaces ,where $\mathcal{M}^+=\{\mathcal{M}^i \in \mathcal{M}|\beta_{m^i \sim \pi_{\mathcal{M}^i}(\cdot|\hat{h}^i_t)} = 1, \, \forall i \in I\}$\\
$t_{max}$ & The maximum execution timestep.\\
$T$& The execution timestep space of a macro-action, where  $T = \{0,\cdots,t_{max}\}$\\

$\boldsymbol{m}^{-}$ & The joint-macro-action over the agents who have not terminated the macro-actions and will continue running next step\\ 
$\boldsymbol{t_m}$ & The joint execution timestep of agents\\
$Q_i$ & The individual value function of agent $i$\\
$Q_{total}$ &  The joint value function of agents\\
$A_i$ & The individual advantage function of agent $i$\\
$A_{total}$ &  The joint advantage function of agents\\
$F^\textit{IGM}$ & The class of functions satisfying IGM\\
$F^\textit{Mac-IGM }$ & The class of functions satisfying Mac-IGM\\
$F^\textit{To-Mac-IGM }$ & The class of functions satisfying To-Mac-IGM\\
$F^\textit{Adv-IGM}$ & The class of functions satisfying Adv-IGM\\
$F^\textit{MacAdv-IGM}$ & The class of functions satisfying MacAdv-IGM\\
$F^\textit{To-MacAdv-IGM}$ & The class of functions satisfying To-MacAdv-IGM\\
$e_t$ & The temporal encodings for agents after temporal self-attention
module, where $e_t=\{e_t^i\}_{i=0}^n$  at timestep $t$\\
$e_t^i$& The temporal encodings for agent $i$ after temporal self-attention
module at timestep $t$\\
$Q_{i,t}^{TEM}$& The query matrix of temporal self-attention
module of agent $i$ at timestep $t$\\
$K_{i}^{TEM}$& The key matrix of temporal self-attention
module of agent $i$ at timestep $t$\\
$V_{i}^{TEM}$& The value matrix of temporal self-attention
module of agent $i$ at timestep $t$\\
$W^{TEM}_Q$& The trainable matrices for generating query matrix $Q_{i,t}^{TEM}$\\
$W^{TEM}_K$& The trainable matrices for generating key matrix $K_{i,t}^{TEM}$\\
$W^{TEM}_V$& The trainable matrices for generating value matrix $V_{i,t}^{TEM}$\\
$M^{TEM}$& The causality mask with its first $t$-th entries equal to 1, and remaining entries 0.\\
$d$&\\
$Q_{i,t}$& The query matrix of agent-oriented self-attention
module of agent $i$ at timestep $t$\\
$K_{i,t}$& The key matrix of agent-oriented self-attention
module of agent $i$ at timestep $t$\\
$V_{i,t}$& The value matrix of agent-oriented self-attention
module of agent $i$ at timestep $t$\\
$\boldsymbol{K_{t}}$&The key matrix of agent-oriented self-attention
module of agents at timestep $t$\\
$\boldsymbol{V_{t}}$&The key matrix of agent-oriented self-attention
module of agents at timestep $t$\\
$W_Q$&The trainable matrices for generating query matrix $Q_{i,t}$\\
$W_K$&The trainable matrices for generating query matrix $K_{i,t}$\\
$W_V$&The trainable matrices for generating query matrix $V_{i,t}$\\
$\alpha_t^{ij}$& The attention weight $\alpha_t^{ij}$, which captures the correlation between agents $i$ and $j$ at timestep $t$\\
$\delta$& The hyper-parameter indicated the minimum degree of interdependence\\
$\hat{e_t^i}$& The final encodings for agent $i$ after agent-oriented self-attention module at timestep $t$\\
$\hat{\boldsymbol{e_t}}$& The final encodings for agents after agent-oriented self-attention module at timestep $t$\\
$\partial$& The partial derivation\\
\bottomrule
\end{tabular}
\end{small}
\end{table}

\begin{thebibliography}{31}
\providecommand{\natexlab}[1]{#1}
\providecommand{\url}[1]{\texttt{#1}}
\expandafter\ifx\csname urlstyle\endcsname\relax
  \providecommand{\doi}[1]{doi: #1}\else
  \providecommand{\doi}{doi: \begingroup \urlstyle{rm}\Url}\fi

\bibitem[Amato et~al.(2019)Amato, Konidaris, Kaelbling, and How]{macPOMDP2}
C.~Amato, G.~Konidaris, L.~P. Kaelbling, and J.~P. How.
\newblock Modeling and planning with macro-actions in decentralized pomdps.
\newblock \emph{Journal of Artificial Intelligence Research}, 64:\penalty0 817--859, 2019.

\bibitem[Amato and Vian(2017)]{asy_marl1}
O.~S. A.-M.~A. Amato and C.~L. S. H.~J. Vian.
\newblock J decentralized control of multi-robot partially observable markov decision processes using belief space macro-actions.
\newblock \emph{The International Journal of Robotics Research}, 36\penalty0 (2):\penalty0 231, 2017.

\bibitem[Christopher et~al.(2014)Christopher, George, and Leslie]{macPOMDP1}
A.~Christopher, D.~K. George, and P.~K. Leslie.
\newblock Planning with macro-actions in decentralized pomdps.
\newblock In \emph{International Conference on Autonomous Agents and Multiagent Systems}, 2014.

\bibitem[Chu et~al.(2019)Chu, Wang, Codec{\`a}, and Li]{traffic1}
T.~Chu, J.~Wang, L.~Codec{\`a}, and Z.~Li.
\newblock Multi-agent deep reinforcement learning for large-scale traffic signal control.
\newblock \emph{IEEE transactions on intelligent transportation systems}, 21\penalty0 (3):\penalty0 1086--1095, 2019.

\bibitem[Colas et~al.(2019)Colas, Sigaud, and Oudeyer]{statistical_comparisons}
C.~Colas, O.~Sigaud, and P.-Y. Oudeyer.
\newblock A hitchhiker's guide to statistical comparisons of reinforcement learning algorithms.
\newblock \emph{arXiv preprint arXiv:1904.06979}, 2019.

\bibitem[Gao et~al.(2021)Gao, Shi, Du, Wang, Chen, Lian, Qiu, Han, Wang, Ye, et~al.]{game}
Y.~Gao, B.~Shi, X.~Du, L.~Wang, G.~Chen, Z.~Lian, F.~Qiu, G.~Han, W.~Wang, D.~Ye, et~al.
\newblock Learning diverse policies in moba games via macro-goals.
\newblock \emph{Advances in Neural Information Processing Systems}, 34:\penalty0 16171--16182, 2021.

\bibitem[Gu et~al.(2023)Gu, Kuba, Chen, Du, Yang, Knoll, and Yang]{robot}
S.~Gu, J.~G. Kuba, Y.~Chen, Y.~Du, L.~Yang, A.~Knoll, and Y.~Yang.
\newblock Safe multi-agent reinforcement learning for multi-robot control.
\newblock \emph{Artificial Intelligence}, 319:\penalty0 103905, 2023.

\bibitem[Gupta et~al.(2021)Gupta, Mahajan, Peng, B{\"o}hmer, and Whiteson]{uneven}
T.~Gupta, A.~Mahajan, B.~Peng, W.~B{\"o}hmer, and S.~Whiteson.
\newblock Uneven: Universal value exploration for multi-agent reinforcement learning.
\newblock In \emph{International Conference on Machine Learning}, pages 3930--3941. PMLR, 2021.

\bibitem[Horn(1990)]{hadamard}
R.~A. Horn.
\newblock The hadamard product.
\newblock In \emph{Proc. Symp. Appl. Math}, volume~40, pages 87--169, 1990.

\bibitem[Liu and Li(2024)]{robot2}
Y.~Liu and J.~Li.
\newblock Runtime verification-based safe marl for optimized safety policy generation for multi-robot systems.
\newblock \emph{Big Data and Cognitive Computing}, 8\penalty0 (5):\penalty0 49, 2024.

\bibitem[Mahajan et~al.(2019)Mahajan, Rashid, Samvelyan, and Whiteson]{maven}
A.~Mahajan, T.~Rashid, M.~Samvelyan, and S.~Whiteson.
\newblock Maven: Multi-agent variational exploration.
\newblock \emph{Advances in neural information processing systems}, 32, 2019.

\bibitem[Marchesini et~al.(2024)Marchesini, Xiao, and Amato]{avf}
E.~Marchesini, Y.~Xiao, and C.~Amato.
\newblock Value factorization for asynchronous multi-agent reinforcement learning.
\newblock \emph{Journal Placeholder}, 2024.

\bibitem[Oliehoek et~al.(2016)Oliehoek, Amato, et~al.]{dec-pomdp}
F.~A. Oliehoek, C.~Amato, et~al.
\newblock \emph{A concise introduction to decentralized POMDPs}, volume~1.
\newblock Springer, 2016.

\bibitem[Omidshafiei et~al.(2017)Omidshafiei, Agha-Mohammadi, Amato, Liu, How, and Vian]{asy_marl2}
S.~Omidshafiei, A.-A. Agha-Mohammadi, C.~Amato, S.-Y. Liu, J.~P. How, and J.~Vian.
\newblock Decentralized control of multi-robot partially observable markov decision processes using belief space macro-actions.
\newblock \emph{The International Journal of Robotics Research}, 36\penalty0 (2):\penalty0 231--258, 2017.

\bibitem[Rashid et~al.(2020{\natexlab{a}})Rashid, Farquhar, Peng, and Whiteson]{wqmix}
T.~Rashid, G.~Farquhar, B.~Peng, and S.~Whiteson.
\newblock Weighted qmix: Expanding monotonic value function factorisation for deep multi-agent reinforcement learning.
\newblock \emph{Advances in neural information processing systems}, 33:\penalty0 10199--10210, 2020{\natexlab{a}}.

\bibitem[Rashid et~al.(2020{\natexlab{b}})Rashid, Samvelyan, De~Witt, Farquhar, Foerster, and Whiteson]{qmix}
T.~Rashid, M.~Samvelyan, C.~S. De~Witt, G.~Farquhar, J.~Foerster, and S.~Whiteson.
\newblock Monotonic value function factorisation for deep multi-agent reinforcement learning.
\newblock \emph{Journal of Machine Learning Research}, 21\penalty0 (178):\penalty0 1--51, 2020{\natexlab{b}}.

\bibitem[Seuken and Zilberstein(2012)]{boxpushing}
S.~Seuken and S.~Zilberstein.
\newblock Improved memory-bounded dynamic programming for decentralized pomdps.
\newblock \emph{arXiv preprint arXiv:1206.5295}, 2012.

\bibitem[Son et~al.(2019)Son, Kim, Kang, Hostallero, and Yi]{qtran}
K.~Son, D.~Kim, W.~J. Kang, D.~E. Hostallero, and Y.~Yi.
\newblock Qtran: Learning to factorize with transformation for cooperative multi-agent reinforcement learning.
\newblock In \emph{International conference on machine learning}, pages 5887--5896. PMLR, 2019.

\bibitem[Tuyls and Weiss(2012)]{ctde}
K.~Tuyls and G.~Weiss.
\newblock Multiagent learning: Basics, challenges, and prospects.
\newblock \emph{Ai Magazine}, 33\penalty0 (3):\penalty0 41--41, 2012.

\bibitem[Wan et~al.(2021)Wan, Liu, Chen, Wang, and Lan]{suboptimal}
L.~Wan, Z.~Liu, X.~Chen, H.~Wang, and X.~Lan.
\newblock Greedy-based value representation for optimal coordination in multi-agent reinforcement learning.
\newblock \emph{arXiv preprint arXiv:2112.04454}, 2021.

\bibitem[Wang et~al.(2020{\natexlab{a}})Wang, Ren, Liu, Yu, and Zhang]{qplex}
J.~Wang, Z.~Ren, T.~Liu, Y.~Yu, and C.~Zhang.
\newblock Qplex: Duplex dueling multi-agent q-learning.
\newblock \emph{arXiv preprint arXiv:2008.01062}, 2020{\natexlab{a}}.

\bibitem[Wang et~al.(2020{\natexlab{b}})Wang, Ke, Qiao, and Chai]{traffic2}
X.~Wang, L.~Ke, Z.~Qiao, and X.~Chai.
\newblock Large-scale traffic signal control using a novel multiagent reinforcement learning.
\newblock \emph{IEEE transactions on cybernetics}, 51\penalty0 (1):\penalty0 174--187, 2020{\natexlab{b}}.

\bibitem[Wu et~al.(2020)Wu, Wang, Evans, Tenenbaum, Parkes, and Kleiman-Weiner]{overcooked}
S.~A. Wu, R.~E. Wang, J.~A. Evans, J.~B. Tenenbaum, D.~C. Parkes, and M.~Kleiman-Weiner.
\newblock Too many cooks: Coordinating multi-agent collaboration through inverse planning.
\newblock In \emph{Proceedings of the annual meeting of the cognitive science society}, volume~42, 2020.

\bibitem[Xiao(2022)]{macros}
Y.~Xiao.
\newblock \emph{Macro-Action-Based Multi-Agent/Robot Deep Reinforcement Learning under Partial Observability}.
\newblock PhD thesis, Northeastern University, 2022.

\bibitem[Xiao et~al.(2020{\natexlab{a}})Xiao, Hoffman, and Amato]{warehouse1}
Y.~Xiao, J.~Hoffman, and C.~Amato.
\newblock Macro-action-based deep multi-agent reinforcement learning.
\newblock In \emph{Conference on Robot Learning}, pages 1146--1161. PMLR, 2020{\natexlab{a}}.

\bibitem[Xiao et~al.(2020{\natexlab{b}})Xiao, Hoffman, Xia, and Amato]{warehouse2}
Y.~Xiao, J.~Hoffman, T.~Xia, and C.~Amato.
\newblock Learning multi-robot decentralized macro-action-based policies via a centralized q-net.
\newblock In \emph{2020 IEEE International conference on robotics and automation (ICRA)}, pages 10695--10701. IEEE, 2020{\natexlab{b}}.

\bibitem[Xiao et~al.(2022)Xiao, Tan, and Amato]{asy_marl3}
Y.~Xiao, W.~Tan, and C.~Amato.
\newblock Asynchronous actor-critic for multi-agent reinforcement learning.
\newblock \emph{Advances in Neural Information Processing Systems}, 35:\penalty0 4385--4400, 2022.

\bibitem[Yang et~al.(2020)Yang, Hao, Liao, Shao, Chen, Liu, and Tang]{qatten}
Y.~Yang, J.~Hao, B.~Liao, K.~Shao, G.~Chen, W.~Liu, and H.~Tang.
\newblock Qatten: A general framework for cooperative multiagent reinforcement learning.
\newblock \emph{arXiv preprint arXiv:2002.03939}, 2020.

\bibitem[Yao et~al.(2021)Yao, Wen, Wang, and Tan]{smix}
X.~Yao, C.~Wen, Y.~Wang, and X.~Tan.
\newblock Smix ($\lambda$): Enhancing centralized value functions for cooperative multiagent reinforcement learning.
\newblock \emph{IEEE Transactions on Neural Networks and Learning Systems}, 34\penalty0 (1):\penalty0 52--63, 2021.

\bibitem[Zhang et~al.(2019)Zhang, Li, Zhang, Zheng, Zhang, Hao, Chen, Chen, Xiao, and Zhou]{game2}
Z.~Zhang, H.~Li, L.~Zhang, T.~Zheng, T.~Zhang, X.~Hao, X.~Chen, M.~Chen, F.~Xiao, and W.~Zhou.
\newblock Hierarchical reinforcement learning for multi-agent moba game.
\newblock \emph{arXiv preprint arXiv:1901.08004}, 2019.

\bibitem[Zhou et~al.(2020)Zhou, Liu, Sui, Li, and Chung]{implicit}
M.~Zhou, Z.~Liu, P.~Sui, Y.~Li, and Y.~Y. Chung.
\newblock Learning implicit credit assignment for cooperative multi-agent reinforcement learning.
\newblock \emph{Advances in neural information processing systems}, 33:\penalty0 11853--11864, 2020.

\end{thebibliography}
\end{document}